\documentclass{iopart}
\usepackage{iopams}
\usepackage{amssymb,amstext}
\usepackage{graphicx}

\begin{document}

\title[Molecular Dynamics study of molten and glassy germanium dioxide]
{Comparative classical and ``ab initio'' Molecular Dynamics study of molten 
and glassy germanium dioxide}
\author{M Hawlitzky$^{1}$, J Horbach$^{1,2}$, S Ispas$^{3}$, M Krack$^{4}$ and
K Binder$^{1}$}
\address{$^{1}$Institut f\"ur Physik, Johannes Gutenberg-Universit\"at Mainz, 
               Staudinger Weg 7, 55099 Mainz, Germany\\
         $^{2}$Institut f\"ur Materialphysik im Weltraum, Deutsches Zentrum f\"ur
               Luft- und Raumfahrt (DLR), 51170 K\"oln, Germany\\
         $^{3}$Laboratoire des Collo\"\i des, Verres et Nanomat\'eriaux, Universit\'e 
               Montpellier II and CNRS UMR 5587, 34095 Montpellier, France\\
         $^{4}$Computational Science, Dept.~of Chemistry and Applied Biosciences,
               ETH Z\"urich, USI Campus, Via Giuseppe Buffi 13, 6900 Lugano, Switzerland}
\eads{\mailto{kurt.binder@uni-mainz.de and juergen.horbach@dlr.de}}
\date{\today}

\begin{abstract}
A Molecular Dynamics (MD) study of static and dynamic properties of
molten and glassy germanium dioxide is presented. The interactions
between the atoms are modelled by the classical pair potential proposed
by Oeffner and Elliott (OE) [Oeffner R D and Elliott S R 1998 {\it
Phys. Rev. B} {\bf 58} 14791]. We compare our results to experiments
and previous simulations. In addition, an ``ab initio'' method, the
so-called Car-Parrinello Molecular Dynamics (CPMD), is applied to check
the accuracy of the structural properties, as obtained by the classical
MD simulations with the OE potential. As in a similar study for SiO$_2$,
the structure predicted by CPMD is only slightly softer than that
resulting from the classical MD. In contrast to earlier simulations,
both the static structure and dynamic properties are in very good
agreement with pertinent experimental data. MD simulations with the OE
potential are also used to study the relaxation dynamics. As previously
found for SiO$_2$, for high temperatures the dynamics of molten GeO$_2$
is compatible with a description in terms of mode coupling theory.

\end{abstract}

%\maketitle

%
\section{Introduction}
Understanding the structure and dynamics of glassforming fluids and the
nature of the glass transition is one of the most challenging unsolved
problems of the physics of condensed matter \cite{1,2,3,4,5,6,7}. One
of the most debated issues is the question to which extent the glass
transition is a universal phenomenon; i.e.~it is debated whether the
mechanisms causing the dramatic slowing down in undercooled fluids
when the glass transition is approached are basically the same in all
glassforming materials, or whether qualitatively different classes
of glass transitions exist, similar to the ``universality classes''
of critical phenomena \cite{8}.

One such distinction in two classes has been proposed by Angell
\cite{9}, namely the distinction between ``strong'' and ``fragile''
glassformers. Plotting the logarithm of the viscosity $\eta (T)$ versus
the normalized inverse temperature $T_{\rm g}/T$ (the glass transition
temperature $T_{\rm g}$ is here defined somewhat arbitrarily from
the condition $\eta (T=T_{\rm g})=10^{13}$\,Poise), one observes that
certain network-forming materials such as molten SiO$_2$ and molten
GeO$_2$ simply follow straight lines, i.e.~the temperature dependence
of $\eta(T)$ can be described by an Arrhenius law,
\begin{equation} \label{eq1}
\eta(T) = \eta_\infty \exp\left(E_{\rm a} /k_BT \right) \; ,
\end{equation}
where $\eta_\infty$ is a constant and $E_{\rm a}$ plays the role of
an activation energy. Most other glassforming systems, however, in
particular polymer melts, multicomponent metallic melts, and fluids
formed from small organic molecules, behave differently. For these
glassformers, the plot of ${\rm log}[\eta(T)]$ vs.~$T_{\rm g}/T$ is
strongly curved. Following Angell \cite{9}, these systems are called
``fragile glassformers''.

There is ample evidence \cite{6,10} that in fragile glassformers
the initial stages of slowing down, when the structural relaxation
times grow from the picosecond scale by several orders of magnitude,
can be described rather well by mode coupling theory (MCT) \cite{3},
although some aspects of this theory are still under discussion
\cite{11}, and there is no consensus on the behavior near $T_{\rm
g}$ \cite{6,12,13,14}. For the case of silica, computer simulation
studies \cite{23,27,30} have shown that the relaxation dynamics at high
temperatures can be well described by MCT, whereas at low temperatures an
Arrhenius behavior is observed, as seen in experiments (note that the
high temperature regime is almost not accessible by experiments). This
indicates that, at least on a qualitative level, the ``strong
glassformers'' SiO$_2$ exhibits a similar behavior for the temperature
dependence of transport coefficients and structural relaxation as
typical ``fragile glassformers''. Now, the question arises whether
this is also true for the other prototype of a ``strong glassformer'',
namely GeO$_2$. While molten silica has been studied extensively,
both by various experimental techniques and by computer simulations
\cite{23,27,30,15,16,17,18,19,20,21,22,24,25,26,28,29,31,32,33,34},
studies of molten and glassy germanium dioxide are
less abundant, and this holds true for both experiments
\cite{35,36,37,38,39,40,41,42,43,salmon06,salmon07_1,salmon07_2} and
simulations \cite{44,45,46,47,48,49,50,giaccomazzi06,micoulaut06_2}.

In the present work, we hence present a detailed Molecular Dynamics
(MD) \cite{51,52} study of molten and glassy GeO$_2$ at zero pressure,
using a pair potential model that has been recently proposed by Oeffner
and Elliott \cite{44}. In order to check whether the Oeffner-Elliott (OE) potential
provides a chemically realistic modelling of GeO$_2$, we perform also
``ab initio'' Car-Parrinello Molecular Dynamics (CPMD) simulations
\cite{53,54,55} and compare various structural and dynamic quantities as
obtained from classical MD using the OE potential with those from the
CPMD calculations. Moreover, our simulation results are also validated
by comparison to experimental data.

In Sec.~2 we summarize the models and methods of the simulation, while
Sec.~3 is devoted to a description of the static properties of molten
and glassy GeO$_2$ (partial pair correlations and structure factors,
ring statistics and angular distributions, etc.). Section~4 presents
selected information on dynamic properties (mean square displacements,
intermediate incoherent scattering functions), while Sec.~5 summarizes
some conclusions.

\section{Models and simulation methods}
\subsection{Classical MD}

In a classical MD simulation, all degrees of freedom due to the
electrons are disregarded, as well as quantum effects due to the ions
(which need to be included for a correct description of thermal
properties of glasses at temperatures far below the glass transition
temperature). One simply solves Newton's equations of motion, which is
conveniently done applying the velocity form of the Verlet algorithm
\cite{51,52}. Forces are computed using the OE
potential \cite{44},
\begin{equation} \label{eq2}
V_{\alpha \beta} (r_{ij})= \frac{q_\alpha q_\beta e^2}{r_{ij}} +
A_{\alpha \beta} \exp (-B_{\alpha\beta} r_{ij}) + C_{\alpha \beta}
r^{-6}_{ij} \quad \quad \alpha, \beta \in {\rm Ge, O} \; .
\end{equation}
Here, $r_{ij}=|\vec{r}_i-\vec{r}_j|$ is the distance between a pair
of particles at positions $\vec{r}_i$ and $\vec{r}_j$. The first
term on the right hand side of Eq.~(\ref{eq2}) describes
Coulomb interactions, with $e$ the elementary charge and the values
$q_{\rm Ge}=1.5$ and $q_{\rm O}=-0.75$ for the partial charges of
germanium and oxygen ions, respectively \cite{44}. The second and
third term in (\ref{eq2}) form a Buckingham potential and describe the
short-range part of the potential.  The constants $B_{\alpha \beta}$
and $C_{\alpha \beta}$ are \cite{44} $A_{\rm GeO}=208011.52$\,eV,
$B_{\rm GeO}=6.129329$\,\AA$^{-1}$, $C_{\rm GeO}=236.653$\,eV\,\AA$^6$,
$A_{\rm OO}=7693.522$\,eV, $B_{\rm OO} =3.285108$\,\AA$^{-1}$,
and $C_{\rm OO}=131.09$\,eV\,\AA$^6$. The Buckingham terms for
the Ge-Ge interaction are set to zero. The OE potential was derived
from quantum-chemical calculations of GeO$_4$ tetrahedra, using
also experimental data from the $\alpha$-GeO$_2$ crystal structure at
$T=300$\,K as input information. We note that analogous procedures for
the chemically similar case of SiO$_2$ have led to the potential due
to van Beest, Kramer and van Santen (``BKS potential'') \cite{56},
which has proven useful to reproduce a great variety of experimental
results rather accurately \cite{23,19,22,24,28}. Note that for SiO$_2$
the effective charges are different ($q_{\rm Si}=2.4$, $q_{\rm O}=-1.2$),
despite the chemical similarity. Thus, a combination of the OE and
the BKS potential would not be suitable for the description of oxide
melts containing both GeO$_2$ and SiO$_2$, since the O-O interaction
is modelled differently in both cases.

Several other potential models were proposed in the literature
\cite{48,57,58}, but structural properties of liquid GeO$_2$ derived
from these potentials are not in good agreement with experiment, and
hence these potentials were not used in the present study.

While the short-range part of Eq.~(\ref{eq1}) was cut off and shifted
to zero at a distance $r_c=7.5$\,\AA~\cite{59}, the long-range Coulomb
interactions were treated by Ewald summation methods \cite{51,59}. The
equations of motion were solved for systems of $N=1152$ atoms using an
integration time step of 1.23\,fs and periodic boundary conditions in
all three spatial directions. The simulations in the $NpT$ ensemble at
constant zero pressure, $p=0$, yielded linear dimensions $L(T)$ of the
cubic simulation box in the range 26.6\,\AA$\; \leq L(T) \leq\;$28.4\,\AA~for
temperatures in the range 2530\,K$\; \leq T \leq 6100\;$\,K. Pressure
was kept constant using the Andersen barostat \cite{60}.
Constant temperature was realized by coupling the system periodically to a 
stochastic heat bath \cite{51}. Note that the
runs in the $NpT$ ensemble were only used to create well equilibrated
initial configurations for runs in the microcanonical $NVE$ ensemble
($V$ denoting the system volume and $E$ its internal energy). Using
force parallelization with message passing interface (MPI) routines,
an efficient use of the J\"ulich multiprocessor system (JUMP) with 32
processors used in parallel was possible. Equilibration times $t_e$
spanned the range from 48.9\,ps (40000 time steps) at $T=6100$\,K
to 11.97\,ns (almost 10$^7$ time steps) at $T=2530$\,K, to generate
8 initial configurations, which then were propagated in the $NVE$
ensemble for the same time interval $t_e$, during which structural
and dynamical properties were recorded. Note that the time $t_e$ was
chosen such that the slower species (Ge) moved on average a distance
of 5.5~\AA~at each temperature. Further implementation details are
documented in Ref.~\cite{59}.

\subsection{CPMD}

A important issue for the CPMD simulations of GeO$_2$ is the quality of
the pseudo potentials, which are a necessary input for the CPMD method
\cite{53,54}. While we found that a pseudo potential due to Goedecker
{\it et al} \cite{62} was computationally too demanding for our purposes,
a pseudo potential based on the general gradient approximation (GGA)
with the BLYP exchange-correlation functional in the Troullier-Martins
parametrization \cite{63} was found to be satisfactory. As an energy
cutoff for the plane waves $E_{\rm cut}= 75$\,Ry was used, similar as in
related work for SiO$_2$ \cite{26}. The time step was 0.0726\,fs. For
the thermostatting of the system, we used Nos\'e-Hoover chains \cite{64}
for each ionic degree of freedom as well as for the electronic degrees
of freedom to counterbalance the energy flow from ions to electrons
\cite{tuckerman94}. The parameters used for the Nos\'e-Hoover chains
can be found in a previous publication \cite{26}.

An important problem in CPMD simulations of amorphous systems is the
generation of suitable initial configurations. While in the case of
SiO$_2$, it was found useful to start from classical MD simulations
using the BKS potential \cite{56} and relax these configurations to
new equilibrium states by CPMD \cite{26,65,66}, in the case of GeO$_2$
(using the OE potential \cite{44}) such a procedure did not converge
\cite{59}. The reason for this failure is that the differences
between equilibrated atomic configurations using either classical MD or CPMD
methods for GeO$_2$ are slightly larger than for SiO$_2$, as far as
interatomic distances, angles etc.~are concerned. At the temperatures
of interest ($T=3760$\,K and $T=3000$\,K), which are far above the
melting temperature $T_m$ of GeO$_2$ ($T_{\rm m}=1389$\,K \cite{67})
it is also too time-consuming to start from a crystalline configuration
and melt it in a CPMD run; thus we decided to start from configurations
generated by classical MD at $T=7000$\,K, where subsequent equilibration
by CPMD turned out to be feasible (for 60 particles this took 53000
CPMD steps, while for 120 particles 21000 CPMD steps were sufficient,
using periodic boundary conditions throughout). Then the temperature
was lowered in a single step to $T=3760$\,K (for $N=60$) or $T=3000$\,K
(for $N=60$ and $N=120$), respectively. 
At $T=3760$\,K, runs over 171000 time steps for equilibration and production
were performed corresponding to a real time of 12.4\,ps. At $T=3000$\,K,
we did runs over 340000 time steps for the system with 60 particles and
420000 time steps for the system with 120 particles, thus covering a time
range of 24.7\,ps and 30.5\,ps, respectively. In order to obtain better
statistics, we averaged over 6 independent simulation runs for each system
size and temperature considered. 

The density was chosen to be
$\rho=3.45$\,g/cm$^{3}$, similar to the equilibrium density resulting
from the classical MD simulations in this temperature range, in order
to be able to compare MD and CPMD results at essentially the same
density. This choice implies linear dimensions of the simulation box of
$L=10.023$\,\AA~for $N=60$ and $L=12.629$\,\AA~for $N=120$. Since the
periodic boundary condition does significantly affect the structure and
correlation functions for distances that exceed $L/2$, the smallness
of $N$ and $L$ clearly is a major disadvantage of our implementation
of CPMD, and prevents us from a meaningful study of intermediate range
order by CPMD. Application of novel versions of ab initio MD, suitable to simulate
significantly larger systems \cite{68}, is desirable, but must be left
to future work.

\begin{figure}
\vspace*{0.2cm}
\begin{center}
\includegraphics[width=0.7\textwidth]{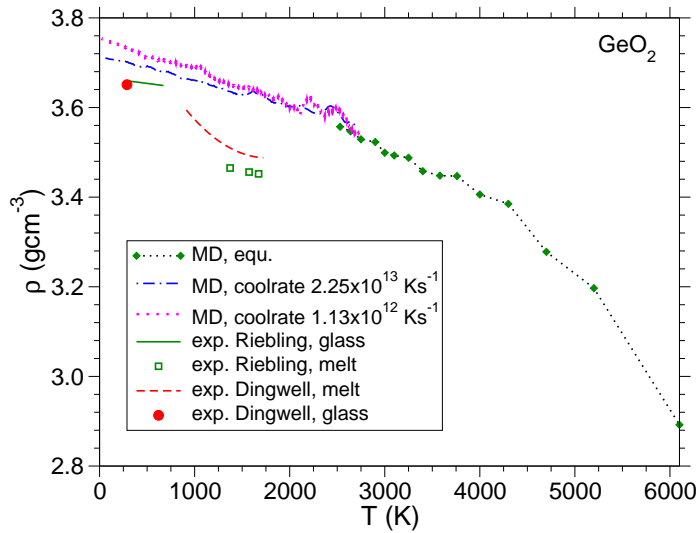}
\caption{\label{fig1}
Density of GeO$_2$ plotted vs.~temperature. Well-equilibrated MD results
(diamonds), using the OE potential, are shown in the temperature range
6100\,K$\ge T \ge$2530\,K. The dotted line connecting the data points
serves only as a guide to the eye. The MD data shown for $T<2750$\,K result
from cooling runs with two different cooling rates, using well-equilibrated
configurations at $T=2530$\,K as a starting point [cf.~Eq.~(\ref{eq3})].
All the simulation results were obtained at zero pressure.  Experimental 
data from Riebling \cite{35} and Dingwell {\it et al} \cite{38} are shown 
for comparison.}
\end{center}
\end{figure}
Finally, we mention that sometimes the generated configurations had to
be discarded ``by hand'', when they contained well-identifiable O$_2$
molecules disjunct from the remaining germanium oxide network (which then
necessarily has coordination defects, of course). It is clear that at
$T=7000$\,K such chemical disintegration of GeO$_2$ may be a physically
meaningful effect. But we are interested in the properties of GeO$_2$
at lower temperatures, where these separate O$_2$ molecules should
no longer occur, but rather should be integrated into the network
structure again. Even at temperatures around 3760\,K, these O$_2$
molecules are expected to be unphysical. If at much lower temperature we were
able to equilibrate the samples over many orders of magnitude in time,
we should see automatically that such defects anneal out again. However,
since this is practically impossible, the somewhat biased selection of
physically meaningful configurations had to be made.

With respect to other implementation details, we closely followed the
procedures of Benoit {\it et al} \cite{26} (see also \cite{59}). We
only note that, in our case, the CPU time required for the CPMD is
a factor of 358000 higher than that needed for the classical MD, using the
same multiprocessor system and the same system size for both methods \cite{59}. Therefore,
only a rather restrictive use of CPMD was feasible.

\section{Static properties of molten and glassy GeO$_2$}

As discussed in Sec.~2.1, equilibration was done in the framework
of classical MD using the $NpT$ ensemble which allows to record the
temperature dependence of the density (Fig.~\ref{fig1}). In our MD
simulation, the lowest temperature which could still be equilibrated
with manageable effort was $T=2530$\,K. This temperature corresponds to
almost twice the melting temperature \cite{67}, while experimental data
are only available at much lower temperatures. Therefore, we used
states at $T=2750$\,K for further cooling down the samples (note that the
states at $T=2530$\,K were not available yet when these cooling runs were 
performed). To this end, temperature was linearly decreased according to
\begin{equation} \label{eq3}
T(t)=2750\, {\rm K}-Qt,
\end{equation}
with cooling rates $Q=2.25 \times 10^{13}$\,K/s and $1.13 \times
10^{12}$\,K/s. As in the case of SiO$_2$ \cite{23,19}, the cooling
rates available in MD exceed those of the experiment by many orders
of magnitude, and a meaningful extrapolation to these very small
experimental cooling rates is not possible. Although the presence of a
density maximum (as is known to occur in SiO$_2$ \cite{69}) somewhere
around $T=2000$\,K cannot be excluded, it seems very unlikely that
for slow cooling rates the simulated densities for $T\leq1700$\,K
would decrease enough to match the experimental data. So we attribute
the larger part of the mismatch between simulated and experimental melt
densities to the inadequacy of the OE potential to predict the density
very accurately! However, such a 5\% discrepancy in the density is not
uncommon when classical pair potentials are used.

Surprisingly, at $T=300$\,K the experimental density is $\rho_{\rm
exp} \approx 3.65$\,g/cm$^3$ and the simulated one (with the slowest
of our cooling rates) $\rho_{\rm sim} \approx3.70$\,g/cm$^3$, thus
only 1.37\% higher. However, this good agreement presumably is due
to a lucky cancellation of errors (freezing in a too high density
due to the inaccurate potential, partially compensates for not
reproducing the rapid variation of the density of supercooled GeO$_2$
around $T=1000$\,K due to our by far too fast cooling). This example
again shows that fits or misfits of isolated experimental data points
by simulations are unsuitable to judge the quality of potentials and/or
simulation procedures.

\begin{figure}
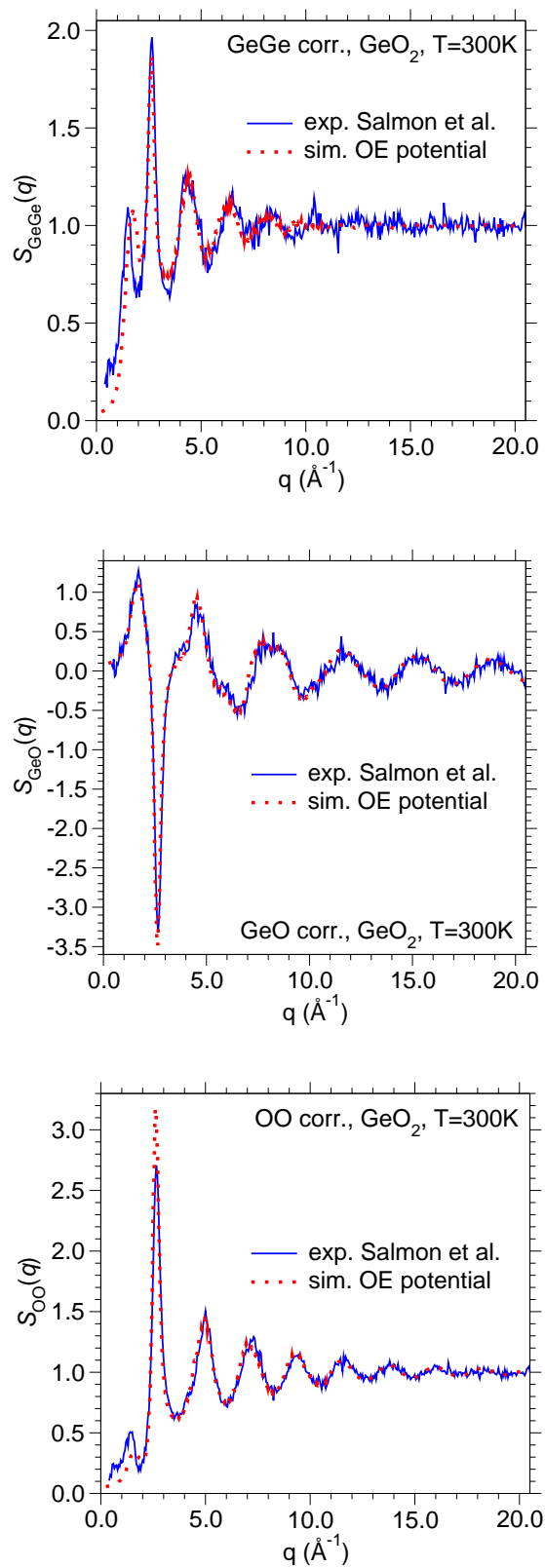

\begin{center}

\vspace*{0.2cm}
\includegraphics[width=0.55\textwidth]{fig2a.eps}
\vspace*{0.7cm}

\includegraphics[width=0.55\textwidth]{fig2b.eps}
\vspace*{0.7cm}

\includegraphics[width=0.55\textwidth]{fig2c.eps}

\caption{\label{fig2}
Partial neutron scattering structure factors $S_{\alpha\beta}(q)$
plotted vs.~wavenumber $q$, comparing the present MD simulation to
the experimental data of Salmon {\it et al} \cite{salmon07_1,salmon07_2} 
at $T=300$\,K.}
\end{center}
\end{figure}
A more detailed information on the static structure, also available via
neutron scattering experiments, is the static structure factor. Since
we deal here with two species, it is appropriate to consider partial
structure factors $S_{\alpha \beta} (q)$ ($\alpha, \beta = {\rm Ge,O}$)
\begin{equation} \label{eq4}
S_{\alpha \beta} (q) = \frac{1}{N} \; \Big\langle
\sum\limits^{N_{\alpha}}_{i=1} \, \,\sum\limits_{j=1}^{N_\beta} \,
\exp (i \vec{q} \cdot \vec{r}_{ij}) \Big\rangle \; .
\end{equation}
Note that fluids and glasses are isotropic and hence $S_{\alpha \beta}
(q)$ depends only on the absolute value $q=|\vec{q}|$ and not on the
direction of the scattering vector $\vec{q}$. Using suitable isotopes,
all partial structure factors for GeO$_2$ have recently been measured
by Salmon {\it et al} \cite{salmon07_1,salmon07_2}. Figure \ref{fig2}
reveals a very good agreement between our simulation results and these data.

\begin{figure}
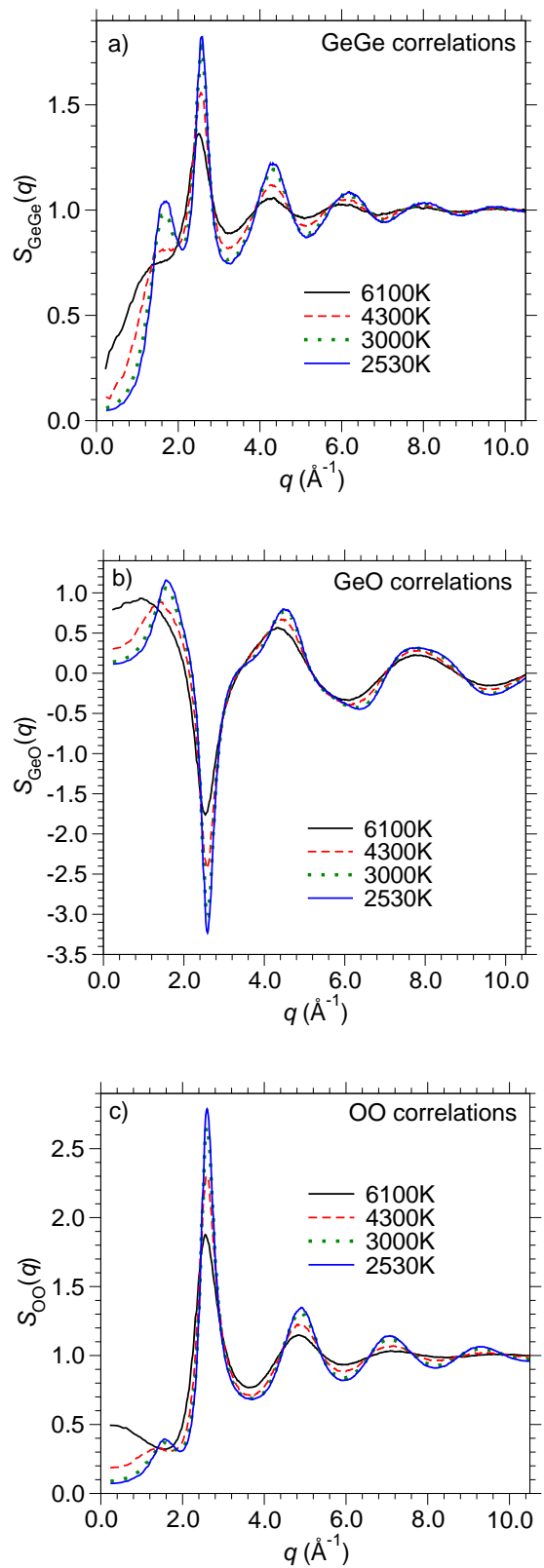

\begin{center}

\vspace*{0.2cm}
\includegraphics[width=0.55\textwidth]{fig3a.eps}
\vspace*{0.7cm}

\includegraphics[width=0.55\textwidth]{fig3b.eps}
\vspace*{0.7cm}

\includegraphics[width=0.55\textwidth]{fig3c.eps}

\caption{\label{fig3} Partial structure factors $S_{\rm GeGe}
(q)$, part a), $S_{\rm GeO} (q)$, part b), and $S_{\rm OO}(q)$,
part c), plotted vs. $q$ and four temperatures, as indicated.}
\end{center}
\end{figure}
Standard neutron scattering yields a scattering intensity weighted with
the scattering lengths $b_{\alpha}$, $b_{\beta}$ as follows
\begin{equation} \label{eq5}
S(q) = \frac{N}{\sum\limits_\alpha \, N_\alpha b^2_\alpha} \,
\sum\limits_{\alpha, \beta \in \{\rm Ge, O\}} b_\alpha b_\beta
\,\, S_{\alpha \beta} (q) \, .
\end{equation}
Using \cite{70} $b_{\rm Ge}=8.185$\,fm, $b_{\rm O}=5.803$\,fm one can
compute from Eqs.~(\ref{eq4},~\ref{eq5}) the neutron scattering structure
factor from the simulation and compare it to corresponding experimental
data \cite{42} without any adjustable parameters whatsoever. Also for
this comparison \cite{47,hawlitzky07} the general agreement between
simulation and experiment is rather good; both predict a ``first sharp
diffraction peak'' (FSDP) \cite{1,6,71} at about $q_{\rm max}\approx
1.55$~\AA$^{-1}$, which can be attributed in real space to the linear
dimension of two GeO$_4$ tetrahedra sharing a corner (see below),
$\ell =2 \pi/q_{\rm max}\approx4.05$\,\AA.

\begin{figure}
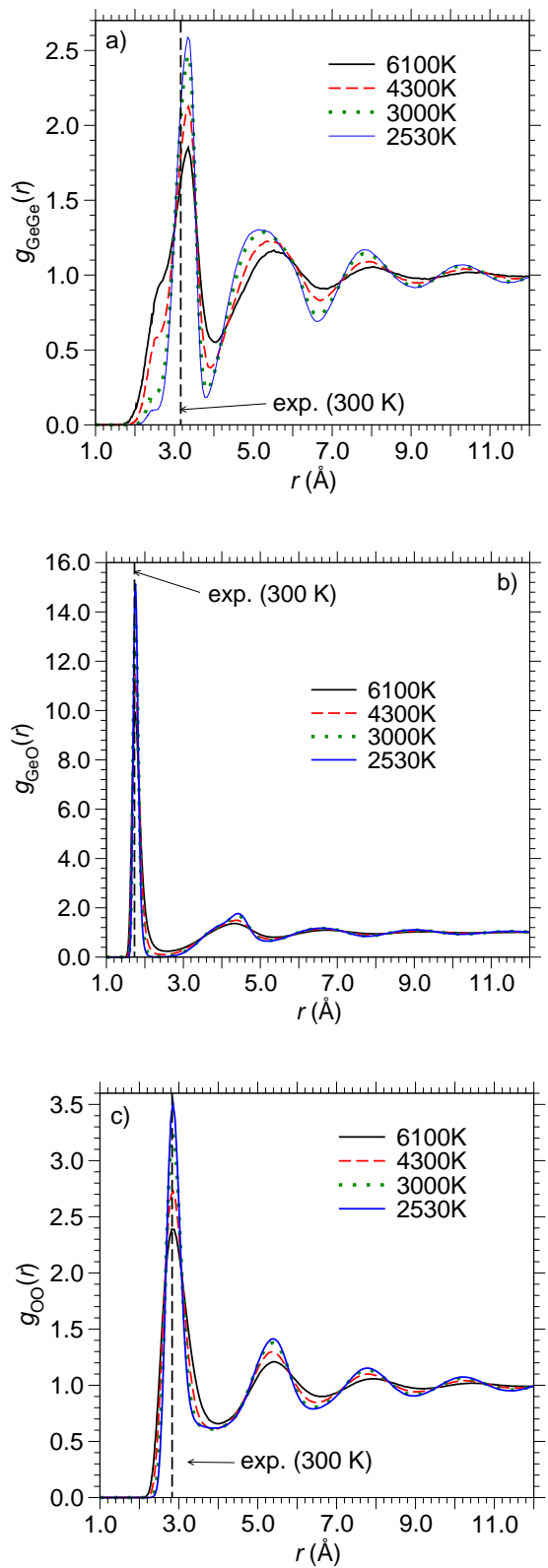

\begin{center}

\vspace*{0.2cm}
\includegraphics[width=0.55\textwidth]{fig4a.eps}
\vspace*{0.7cm}

\includegraphics[width=0.55\textwidth]{fig4b.eps}
\vspace*{0.7cm}

\includegraphics[width=0.55\textwidth]{fig4c.eps}

\caption{\label{fig4} Partial pair correlation functions
$g_{\alpha \beta} (r)$ for GeO$_2$ plotted vs.~$r$ for various
temperatures, as obtained from the classical MD results. The broken
vertical straight line indicates the estimates of nearest neighbor
distances of the Ge-Ge-pairs (a), Ge-O-pairs (b) and O-O pairs (c)
as extracted experimentally from measurements of partial structure
factors at $T=300$\,K \cite{salmon06}.}
\end{center}
\end{figure}
\begin{figure}
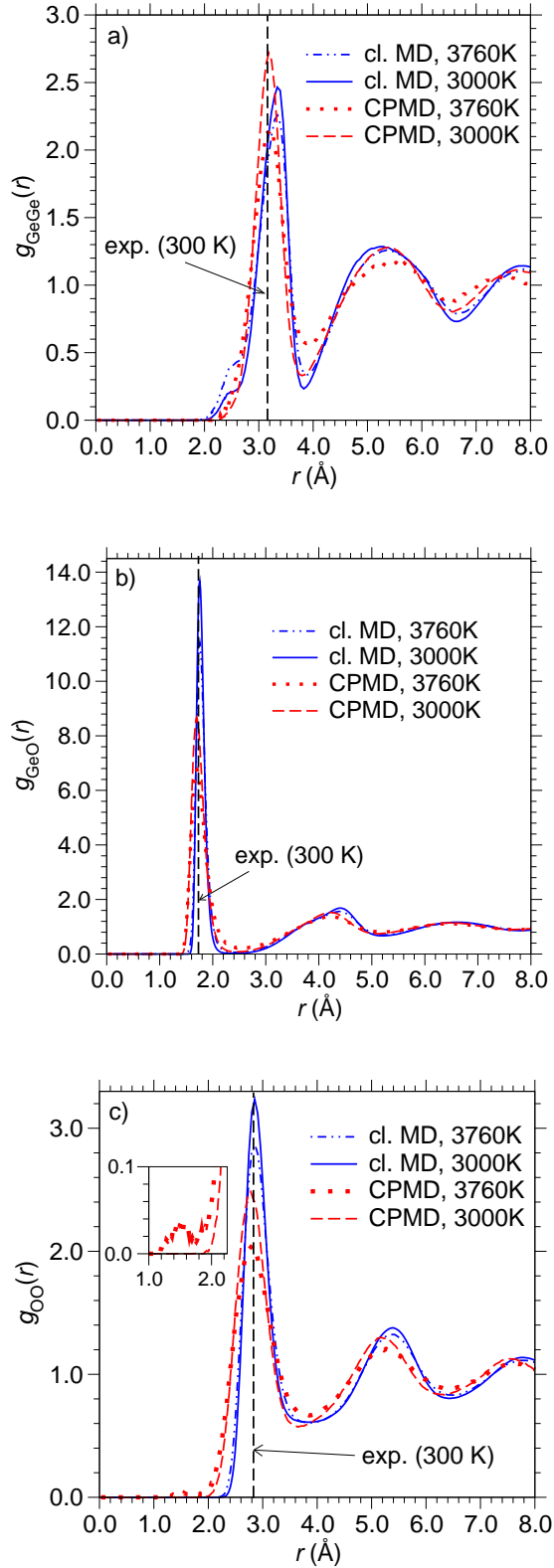

\begin{center}

\vspace*{0.2cm}
\includegraphics[width=0.55\textwidth]{fig5a.eps}
\vspace*{0.7cm}

\includegraphics[width=0.55\textwidth]{fig5b.eps}
\vspace*{0.7cm}

\includegraphics[width=0.55\textwidth]{fig5c.eps}

\caption{\label{fig5} Partial pair correlation functions 
$g_{\alpha \beta (r)}$ for GeO$_2$ plotted $r$ at $T=3000$\,K and 
3760\,K, comparing classical MD with CPMD.  Also the experimental 
estimates for the nearest neighbor distance at $T=300$\,K are 
included \cite{salmon06}. The insert in the oxygen-oxygen
correlation (part c) shows a magnified view of the side maximum
appearing around $r \approx 1.5$\,\AA.}
\end{center}
\end{figure}
When we compare to SiO$_2$ \cite{23,72} we note that in SiO$_2$ the FSDP
occurs at a slightly larger value, $q_{\rm max} \approx 1.7$\,\AA$^{-1}$,
implying a somewhat smaller linear dimension of the two corner-sharing
SiO$_4$ tetrahedra (note that the ``chemical rules'' \cite{1} for
the formation of perfect binary continuous random networks, with a
cation in the center of a tetrahedron and oxygens at the corners,
such that each oxygen is shared by two neighboring tetrahedra, are
identical for SiO$_2$ and GeO$_2$, of course). But a more interesting
difference is the fact that SiO$_2$ shows a second well-developed peak,
at about $q'_{\rm max} \approx 3$\,\AA$^{-1}$, which corresponds to a
peak in GeO$_2$ at about $q'_{\rm max} \approx2.6$\,\AA$^{-1}$. While
in the total neutron scattering structure factor this peak is hardly
distinguishable from the noise, the partial static structure factors
(Figs.~\ref{fig2}, \ref{fig3}) reveal that actually this is the main peak
in the structure, corresponding to a distance $\ell'=2 \pi/q'_{\rm max}
\approx 2.4$\,\AA. This distance, however, cannot be attributed to any
interatomic distance in the structure of GeO$_2$ directly. It rather
corresponds to the period of the oscillatory decay of the partial pair
correlation functions $g_{\alpha \beta}(r)$ in real space at large
distances (Fig.~\ref{fig4}). These correlations are obtained from the
simulated configurations from their definition
\begin{equation} \label{eq6}
g_{\alpha \beta}(r)= N_{\alpha \beta} \Big\langle
\sum\limits_{i=1}^{N _\alpha} \, \sum\limits_{j=1}^{N_\beta} \,
\frac{1}{4 \pi r^2} \delta (r-|\vec{r}_i - \vec{r}_j|) \Big\rangle,
\quad \alpha, \beta = \{{\rm Ge, O}\},
\end{equation}
where $\mathcal{N}_{\alpha, \beta}=V/(N_\alpha N_\beta)$ if $\alpha \neq
\beta$ while $\mathcal{N}_{\alpha \alpha}=V/[\mathcal{N}(N_{\alpha-1})]$,
$V$ being the volume of the simulation box. The correlation functions
$g_{\alpha \beta}(r)$ and $S_{\alpha \beta}(q)$ are related via
\cite{6,73}
\begin{equation} \label{eq7}
S_{\alpha \beta}(q)=1+(N/V) \, \int [g_{\alpha \beta} (\vec{r})-1]
\; \exp (i \vec{q} \cdot \vec{r}) \, d \vec{r} \, .
\end{equation}
In the following, we shall focus on $g_{\alpha \beta}$(r) rather
than on $S_{\alpha \beta}(q)$. Ref.~\cite{salmon06} did give some estimates
of the nearest neighbor distances of the various types of pairs, which
are included in Fig.~\ref{fig4}, indicating a reasonable agreement with the
simulation. Note that the Ge-O distance (about 1.73\,\AA) clearly is
the smallest distance occurring in the structure, and the sharpness of
this peak [note the ordinate scale of Fig.~\ref{fig4}b) in comparison
to that of Fig.~\ref{fig4}a!] reveals that the GeO$_4$ tetrahedra are
fairly rigid. Only for the Ge-Ge distance a slight systematic discrepancy
between MD and experiment is visible. Comparing to the CPMD results
(Fig.~\ref{fig5}), however, this discrepancy seems to be removed.

\begin{figure}
\begin{center}

\includegraphics[width=0.65\textwidth]{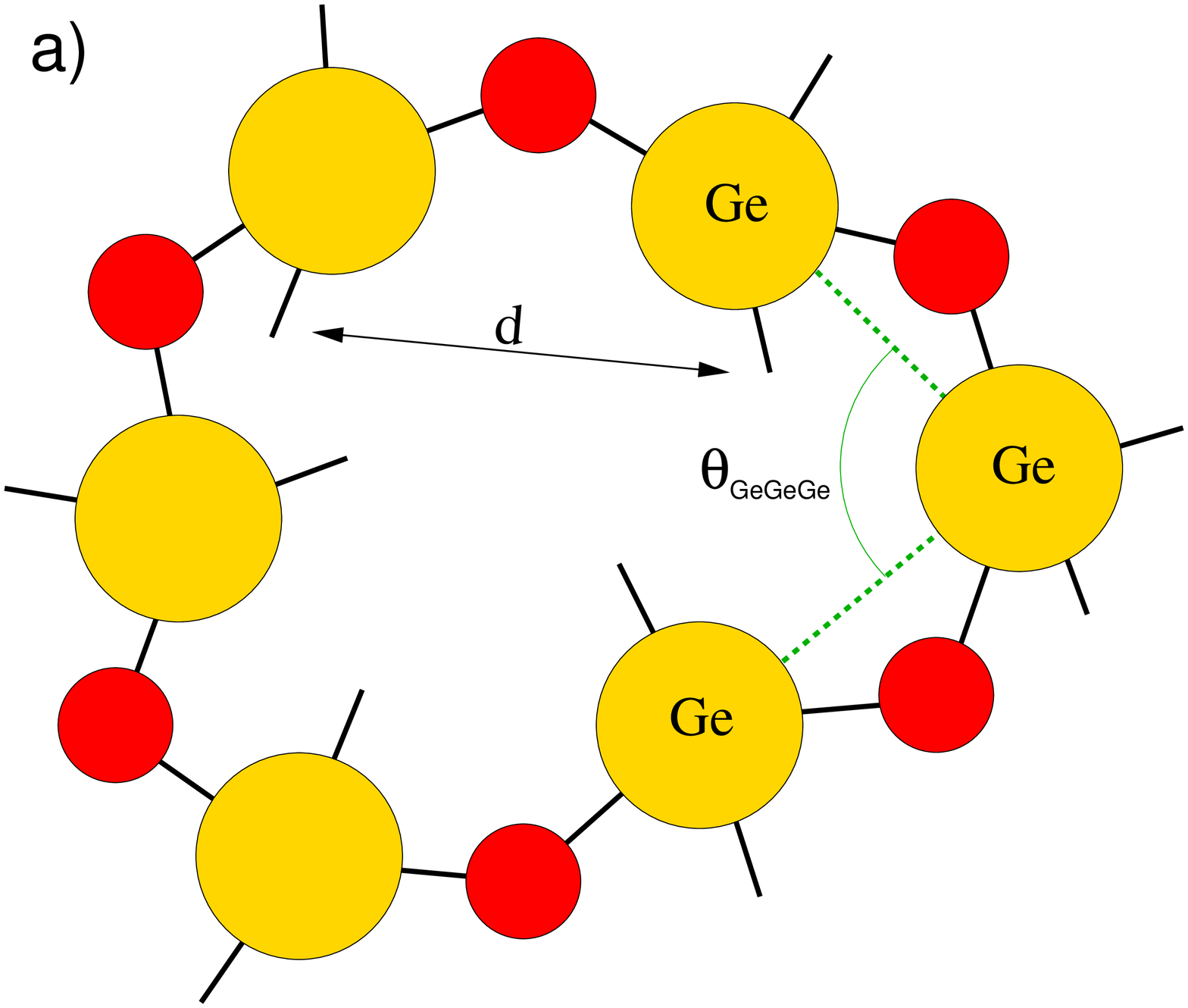}
\vspace*{0.5cm}

\hspace*{1.8cm}
\includegraphics[width=0.35\textwidth]{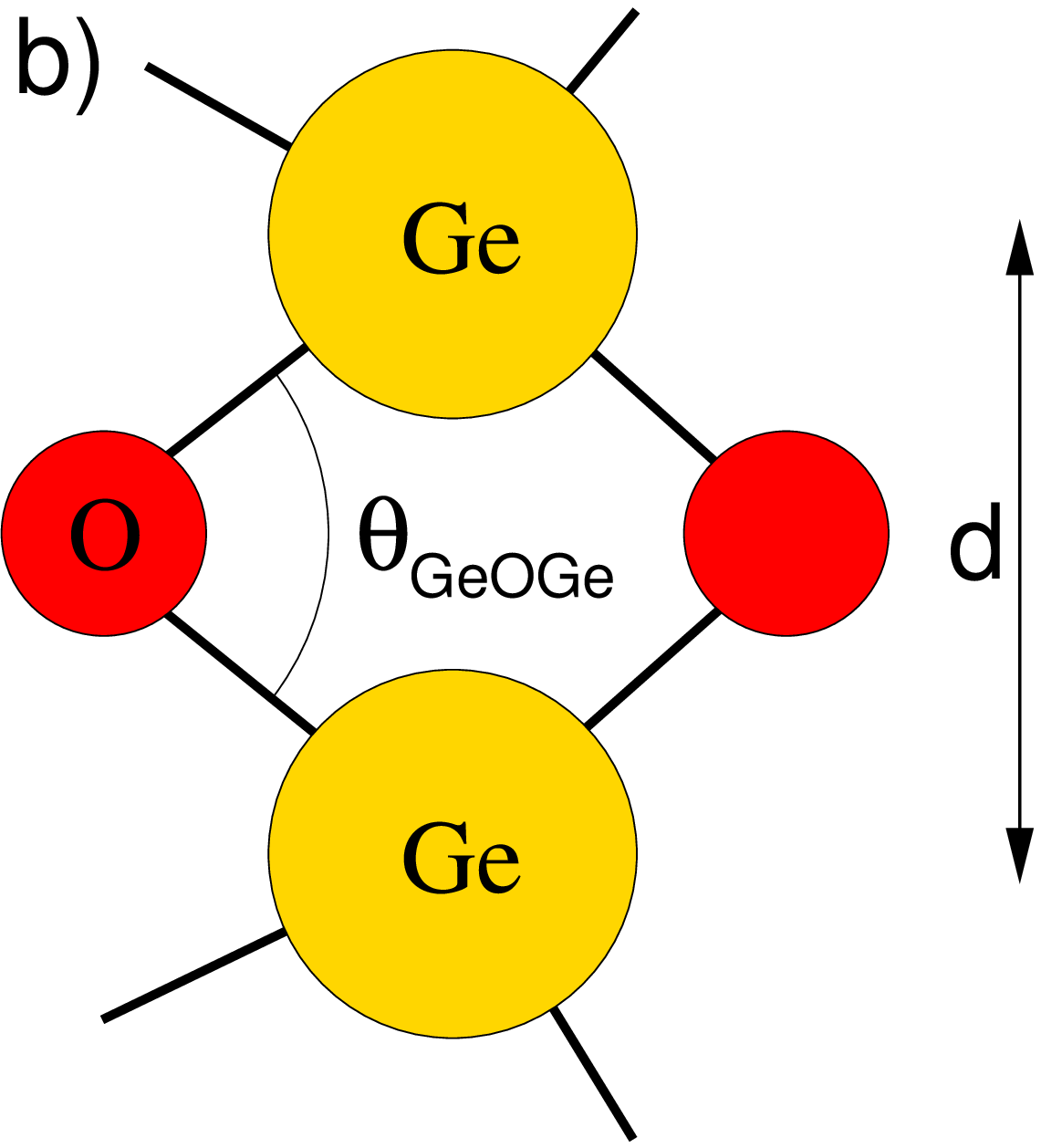}
\hspace*{0.8cm}
%\vspace*{-0.5cm}
\includegraphics[width=0.35\textwidth]{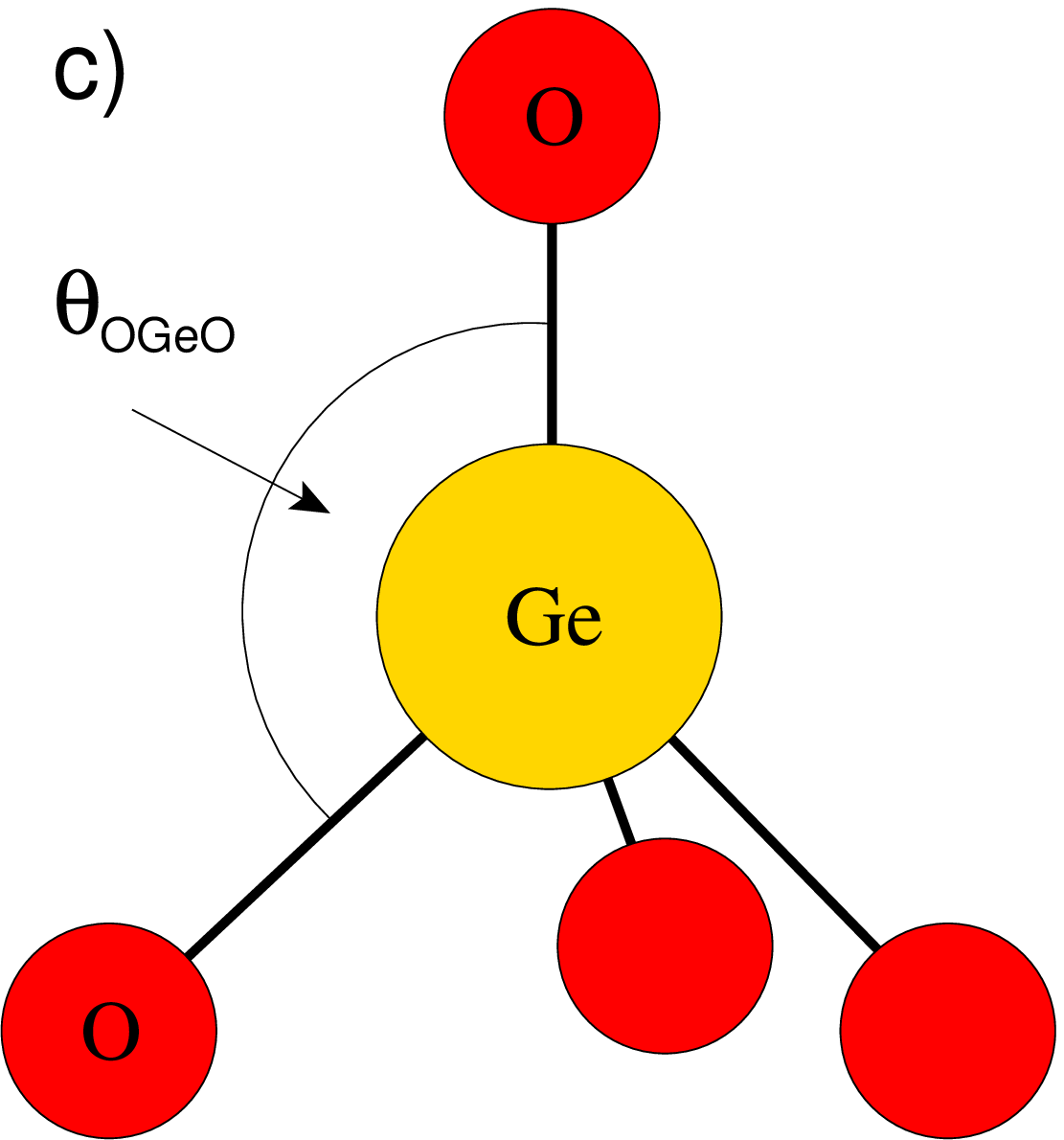}

\caption{\label{fig6} Schematic picture of a ring of length $n=6$,
illustrating also the definition of an angle $\theta_{\rm GeGeGe}$
and the distance $d$ between neighboring Ge atoms (a). A ring of 
length $n=2$ and the angle $\theta_{\rm GeO Ge}$ is sketched in (b), 
and the tetrahedral angle $\theta_{\rm O Ge O}$ in (c).}
\end{center}
\end{figure}
It is also possible to compare CPMD results for $S_{\alpha \beta}(q)$
and $S(q)$ with the corresponding MD results \cite{59}; these comparisons
strengthen the conclusion that one can also draw from Fig.~\ref{fig5},
namely that MD yields a rather accurate description of the local
structure of molten GeO$_2$. Note that slight discrepancies in $g_{\rm OO}(r)$ 
for $r>5$\,\AA~should not be taken very seriously,
because at these distances CPMD suffers from finite size effects, as
noted above. More interesting is the difference (emphasized in the insert
of Fig.~\ref{fig5}c) concerning the feature near 1.5\,\AA. Testing
carefully different equilibration times it was possible to show that
too short equilibration of CPMD yields such prepeak in $g_{\rm OO}(r)$ which is too high
rather than too low \cite{59}. Therefore, this difference between the
CPMD and the MD results is probably a real effect, at least it is not
an artifact of too short equilibration. Of course, one can question
the accuracy of CPMD somewhat on other grounds: other ``ab initio''
studies of the GeO$_2$ structure \cite{47,74} using different
pseudopotentials and system preparation procedures predicted somewhat
different results (e.g.~the Ge-O distance $r_{\rm GeO}=1.69$\,\AA~\cite{74} 
or $r_{{\rm GeO}} =1.78$\,\AA~\cite{47}, while we obtain
1.71\,\AA~and the experimental value is $1.73 \pm 0.03$\,\AA~\cite{39,salmon06}).

\begin{figure}
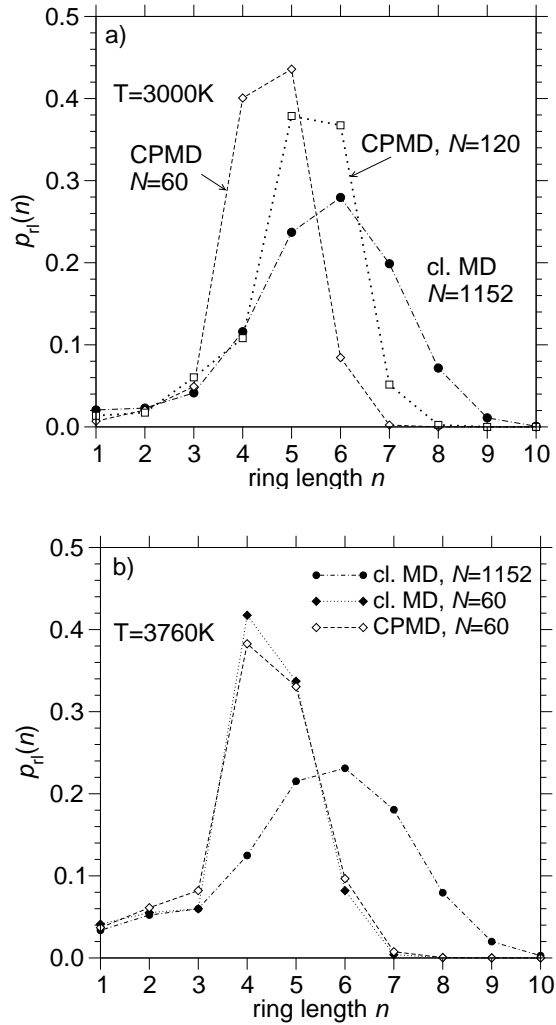

\begin{center}

\vspace*{0.5cm}
\includegraphics[width=0.55\textwidth]{fig7a.eps}
\vspace*{0.5cm}

\includegraphics[width=0.55\textwidth]{fig7b.eps}

\caption{\label{fig7} Probability $P(n)$ that a ring of length $n$
occurs, plotted vs.~$n$, for $T=3000$\,K and $T=3760$\,K, comparing MD
and CPMD (a), and the same comparison including an MD study where
$N=60$, as in the CPMD calculation (b).}
\end{center}
\end{figure}
We now turn our attention to the analysis of structural features
on intermediate length scales. To this end, we recall
the concept of ``ring statistics'' \cite{6,19}. One considers the
shortest closed paths in the network of covalent bonds, starting from
an oxygen atom (Fig.~\ref{fig6}). The length $n$ of a ring is then
the number of cations (Ge in the present case, or Si in the case of
silica \cite{19}) that one passes before one returns to the starting
point. Figure \ref{fig6} shows, as an example, $n=6$ (left) and $n=2$
(middle part). In SiO$_2$, it has been found that the angles between
atoms in a ring with $n=2$ and $n=3$ differ appreciably between the
classical MD simulation and its CPMD counterpart \cite{65,66}, and
this affects also significantly the probability $P(n)$ that a ring of
length $n$ occurs in the structure (in thermal equilibrium). At first
sight one might conclude that a similar effect occurs for GeO$_2$,
too (Fig.~\ref{fig7}a), but a closer analysis reveals that most of the
differences between CPMD and MD stem from the fact that the former
suffers from finite size effects (Fig.~\ref{fig7}b): when we use
$N=60$ in the MD calculation, we find almost perfect agreement with the
CPMD calculation that uses $N=60$ as well. Also the strong difference
between the CPMD results for $N$=60 and $N$=120 show that one cannot
trust the CPMD results for $P(n)$, due to these dominating finite
size effects. Clearly, for a quantity that depends sensitively on the
order on intermediate length scales like $P(n)$ it is more important
to choose a large enough system rather than to work with very realistic
descriptions of the forces, as provided by CPMD.

\begin{figure}
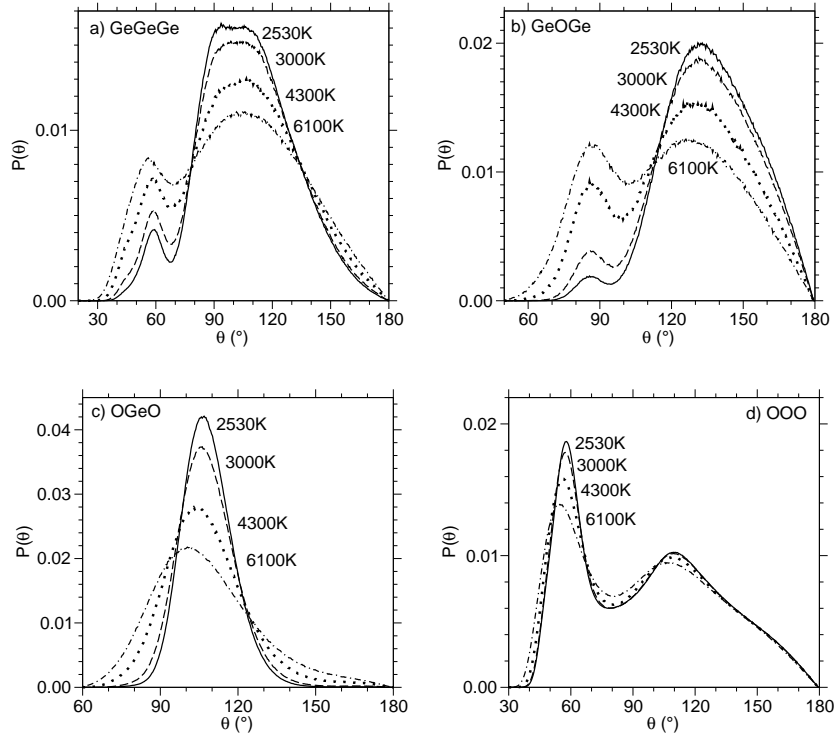

\begin{center}

\vspace*{0.2cm}
\hspace*{1.9cm}
\includegraphics[width=0.4\textwidth]{fig8a.eps}
\hspace*{0.2cm}
\includegraphics[width=0.4\textwidth]{fig8b.eps}
\vspace*{0.5cm}

\hspace*{1.9cm}
\includegraphics[width=0.4\textwidth]{fig8c.eps}
\hspace*{0.2cm}
\includegraphics[width=0.4\textwidth]{fig8d.eps}

\caption{\label{fig8} Distribution functions $P(\theta)$ of
various angles $\theta$, obtained from MD for a wide range of
temperatures, as indicated. Case a) shows the Ge-Ge-Ge angle, case
b) the Ge-O angle, case c) the O-Ge angle and case d) the O-O-O
angle.}
\end{center}
\end{figure}
While CPMD hence is less useful for the study of properties that depend
sensitively on medium range order, it clearly is of great interest
for the assessment of local properties, such as the distributions of
angles between the ``bonds'' in the structure. These distributions
have also been obtained by MD for a wide range of temperatures
(Fig.~\ref{fig8}). The definition of the Ge-Ge-Ge angle is indicated in
Fig.~\ref{fig6}a); other angles are defined analogously. A remarkable
feature is that all distributions, with the exception of the tetrahedral
angle O-Ge-O, have a double peak shape, and are rather broad. Only the
distribution of the tetrahedral angle tends towards a Gaussian shape,
as the temperature is lowered, and gets somewhat sharper; in a random
network structure formed by ideal tetrahedra only, this distribution
would be a delta function, $\delta (\theta-\theta_{\rm tetr})$ with
$\theta_{\rm tetr}=109^\circ$.

The relative weight of the peak at $\theta=60^\circ$ of the Ge-Ge-Ge
angle decreases with decreasing temperature, as well as the weight
of the peak at $\theta=90^\circ$ for the Ge-O-Ge angle distribution. A
consideration of the geometry of the rings (Fig.~\ref{fig6}) immediately
shows that the peak of $P(\theta)$ for the Ge-Ge-Ge angle can be
attributed to rings with $n=3$, and similarly the peak of $P(\theta)$ for
the Ge-O-Ge angle at $\theta=90^\circ$ is due to rings with $n=2$. Such
small rings can be frequently observed in the structure of GeO$_2$ at
high temperatures, while at low temperatures the network becomes much
more regular, and the density of all small rings decreases significantly.

For $T=2530$\,K, the position of the main peak of the distribution
$P(\theta)$ for the Ge-O-Ge angle is 133$^\circ$. It is gratifying
that this number coincides with corresponding experimental estimates
\cite{39,75}. This agreement is a further indication that the OE
potential is able to provide a rather realistic description of the
structure.

The side peaks of Fig.~\ref{fig8}a,b, tend to disappear at the physically
relevant temperatures, i.e.~the number of rings with $n=2$ and $n=3$
becomes significantly smaller with decreasing temperature. This is also
indicated by the temperature dependence of the ring length distribution
$P(n)$ \cite{59}. The main peaks of the GeGeGe and GeOGe distributions
seem to stay rather broad, as expected due to the disorder in the
network structure. Only in the various crystal structures of GeO$_2$
at low temperatures we would expect very sharp distributions of all
angles; in the glass structure only the distributions of the angles
inside a tetrahedron become rather sharp at low temperatures.

In this respect, the distribution of the angle $\theta$ between O-O-O
bonds is special: Fig.~\ref{fig8}d shows that there two peaks occur,
which clearly persist at low temperatures. The obvious explanation is
that there are two distinct possibilities: the peak at $\theta=60^\circ$
can be attributed to oxygen atoms belonging to the same tetrahedron,
while the peak at $\theta\approx 110^\circ$ is due to oxygens belonging
to two neighboring tetrahedra. In fact, as temperature decreases the
structure of a single tetrahedron approaches more and more that of
an ideal tetrahedron, whose faces are perfect triangles, having
angles of 60$^\circ$. In view of this, the observation that the peak
at $\theta=60^\circ$ becomes clearly sharper with decreasing temperature
is not surprising.

\begin{figure}
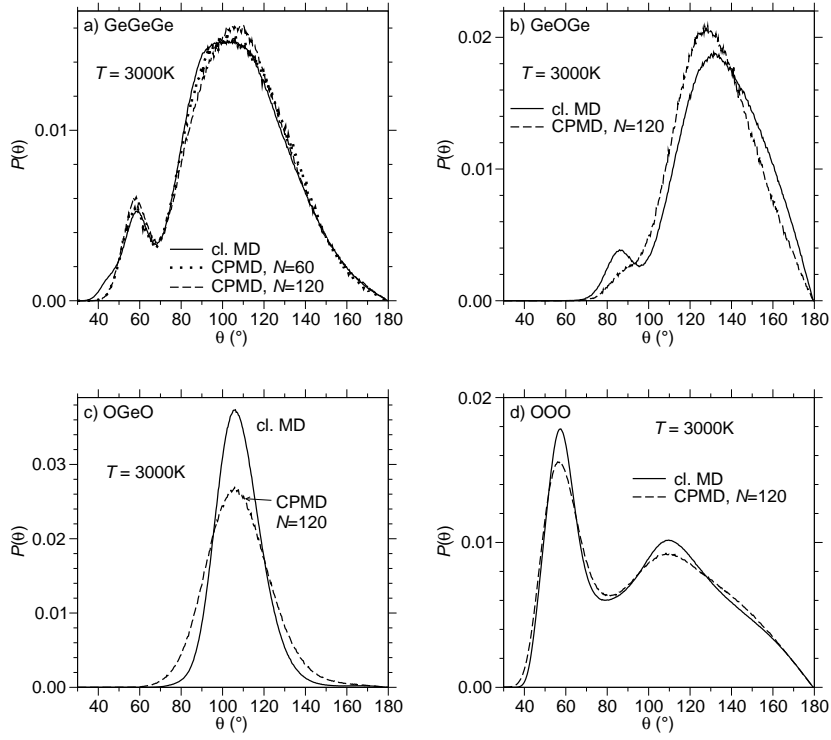


\vspace*{0.2cm}
\hspace*{1.9cm}
\includegraphics[width=0.4\textwidth]{fig9a.eps}
\hspace*{0.2cm}
\includegraphics[width=0.4\textwidth]{fig9b.eps}
\vspace*{0.5cm}

\hspace*{1.9cm}
\includegraphics[width=0.4\textwidth]{fig9c.eps}
\hspace*{0.2cm}
\includegraphics[width=0.4\textwidth]{fig9d.eps}

\caption{\label{fig9} Comparison between MD and CPMD results
at $T=3000$\,K for the distribution functions of various angles, 
Ge-Ge-Ge (a), Ge-O-Ge (b), O-Ge-O (c) and O-O-O (d). All MD results
refer to $N=1152$, while the CPMD are results are for $N=120$ (for 
the Ge-Ge-Ge distribution also the CPMD result for $N=60$ is shown).}
\end{figure}
Now we turn to the comparison of these angular distributions to the
corresponding CPMD predictions (Fig.~\ref{fig9}). The general shape of
these distributions is very similar, with the exception of the Ge-O-Ge
angle, where the side peak at 90$^\circ$ (due to rings with $n=2$)
is broadened into a shoulder only, indicating that the OE potential
overestimates in particular the rigidity of this structural element (a
schematic picture of a ring with $n=2$ is shown in Fig.~\ref{fig6}b)). We
also note that the CPMD distributions are always somewhat broader
than the MD results at the corresponding temperature. This indicates
that the CPMD calculation, if we could parametrize it in terms of an
effective pair potentials having the OE or BKS form, would yield a
systematically softer potential. In fact, if one compares the CPMD
calculation at $T=3000$\,K to the classical calculation at $T=3760$\,K,
the differences are much smaller \cite{59}. Of course, we do not wish to imply
that the differences between CPMD and MD could be fully eliminated by
a renormalization of the temperature scale: for the main peak of the
Ge-O-Ge angle distribution, CPMD at $T=3000$\,K implies a peak at about
129$^\circ$, while the MD calculation yields a peak at about 133$^\circ$
(this value depends much less on temperature than the CPMD peak position
does). We have also done MD simulations with $N=60$ particles only,
to rule out that the differences seen in Fig.~\ref{fig9} simply are
due to finite size effects \cite{59}. Figure \ref{fig9}a also indicates
for the Ge-Ge-Ge distribution that CPMD result for $N=60$ is only
slightly different from that at $N=120$ (for the other distributions
the differences are even smaller).

As a conclusion of this section we may state that the OE potential
predicts slightly too rigid structures in comparison to CPMD, and this
difference is most pronounced at rather high temperatures. However,
the overall agreement between the structure as predicted by the OE
potential and the structure resulting from CPMD is very good. The
same conclusion emerges also from an analysis of the distribution of
coordination numbers \cite{59}. The comparison to experimental data,
whenever available, also suggests the statement that the OE potential
provides a reasonably accurate description of the static structure of
molten and glassy GeO$_2$.

\section{Dynamic properties of GeO$_2$ melts}
\begin{figure}
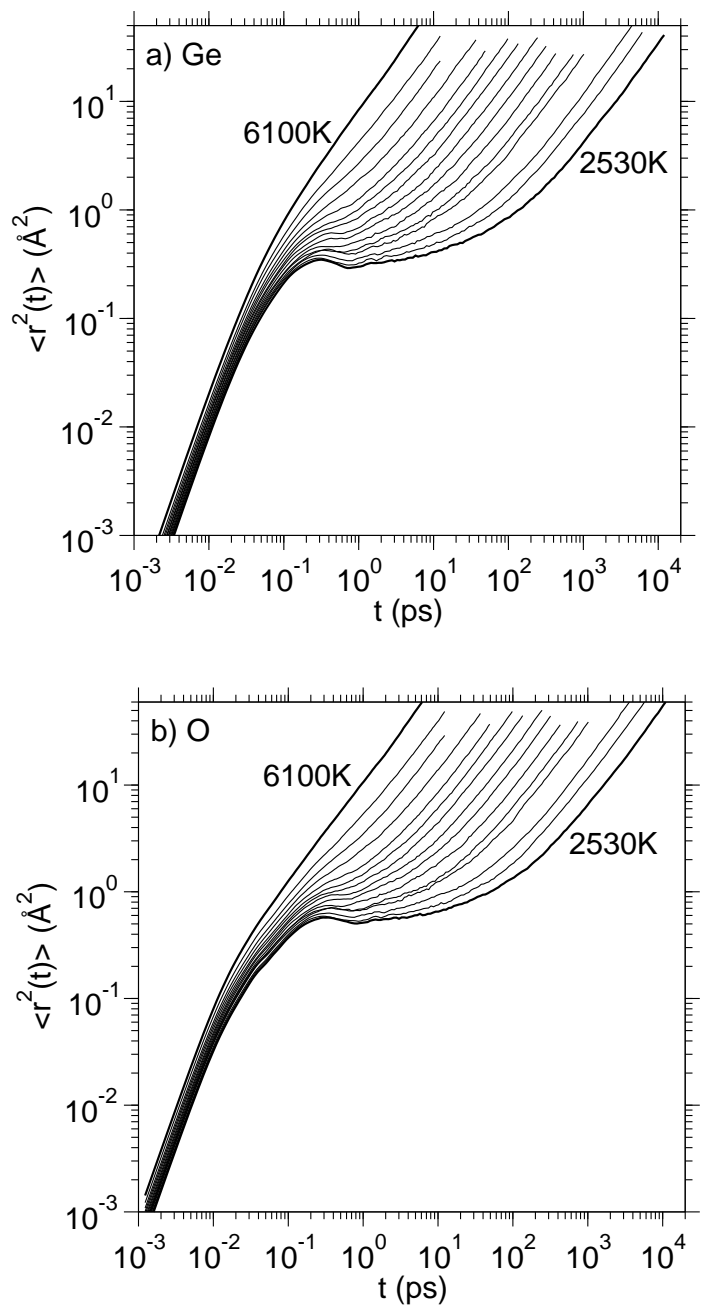

\begin{center}

\vspace*{0.5cm}
\includegraphics[width=0.7\textwidth]{fig10a.eps}
\vspace*{0.7cm}

\includegraphics[width=0.7\textwidth]{fig10b.eps}

\caption{\label{fig10} Log-log plot of the MSD for Ge(a) and O(b)
versus time, for temperatures ranging from $T=2530$\,K
to $T=6100$\,K.}
\end{center}
\end{figure}
From the MD runs in the $NVE$ ensemble, it is straightforward to record
both the mean square displacements (MSD) of a tagged particle of type
$\alpha$ ($\alpha = \{{\rm Ge,O}\}$)
 \cite{4,6,73},
\begin{equation} \label{eq8}
\Big\langle r^2_{\alpha} (t) \Big\rangle =
\frac{1}{N_\alpha} \,
\sum\limits^{N_\alpha}_{i=1} \, 
\Big\langle|\vec{r}_i (t) -
\vec{r}_i(0)|^2 \Big\rangle,
\end{equation}
and the intermediate incoherent scattering function
\begin{equation} \label{eq9}
F^\alpha_{\rm s} (q,t) =
\frac{1}{N_\alpha} \, \sum\limits^{N_\alpha}_{i=1}
\Big\langle \exp \{-i \vec{q} \cdot [\vec{r}_i (t) - \vec{r}_i (0)]\}
\Big\rangle \, .
\end{equation}
The MSD allows to estimate the self-diffusion constants,
applying the Einstein relation
\begin{equation} \label{eq10}
D_\alpha =\lim\limits_{t \rightarrow \infty} 
\left[ \Big\langle r^2_\alpha (t) \Big\rangle /(6t) \right] .
\end{equation}

Figure~\ref{fig10} shows our MD data for the MSD. One sees the
standard behavior, familiar from MD simulations for SiO$_2$ \cite{23}
and many other systems \cite{6}. At very short times, a ballistic
regime is seen ($\langle r^2_{\alpha} (t) \rangle \propto t^2$). Then,
at high temperatures, a rapid crossover to the linear diffusive regime
occurs ($\langle r^2_{\alpha} (t) \rangle = 6D_\alpha t$), while at
lower temperatures, i.e.~in the range 2530\,K$\leq T \leq 3250$\,K,
a plateau is observed at intermediate times, where the MSD does not
increase, but rather stays constant at about $\langle r^2_{\alpha}
(t) \rangle \approx0.5$\,\AA$^2$. This plateau commonly is interpreted
as the onset of the ``cage effect'' \cite{3,6}: each atom sits in a
``cage'' formed by its nearest neighbors and the lower the temperature
the more time it takes until the atom can ``escape from the cage''. Of
course, such mobility implies that the network of bonds in the random
network structure is not rigid, sometimes a bond ``breaks'' \cite{23}
and coordination defects appear, which later can anneal again.

\begin{figure}
\begin{center}
\vspace*{0.5cm}
\includegraphics[width=0.7\textwidth]{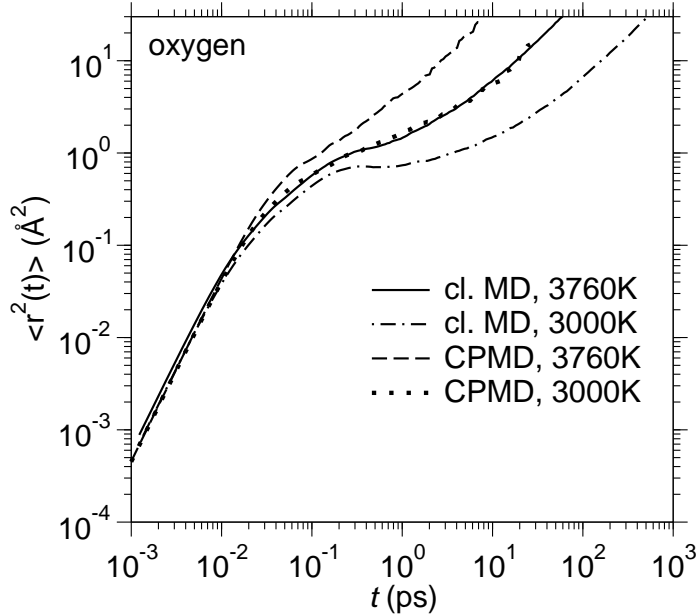}

\caption{\label{fig11} Log-log plot of the MSD for Ge(a) and O(b)
versus time, comparing MD and CPMD results at two temperatures,
$T=3000$\,K and $T=3760$\,K.}
\end{center}
\end{figure}
The MSD, as obtained from MD simulation with the OE model, can be
also compared to corresponding CPMD results. Figure~\ref{fig11}
shows a behavior which is not surprising at all, in view of our
findings for static properties as described in detail in the previous
section: The time dependence of the MSD found for $T=3000$\,K by
CPMD superimposes almost exactly with the MD results for $T=3760$\,K,
reflecting again the finding that CPMD is essentially equivalent to the
use of pair potentials that are slightly softer than the OE potential
but otherwise very similar. As a further caveat we mention the effect
of the Nos\'e-Hoover thermostat (needed in CPMD, not in MD), which may
have speeded up slightly the CPMD dynamics, though we do not have any
real evidence that this effect is already important on time scales up
to 20\,ps that are shown in Fig.~\ref{fig11}.

\begin{figure}
\begin{center}
\vspace*{0.5cm}
\includegraphics[width=0.7\textwidth]{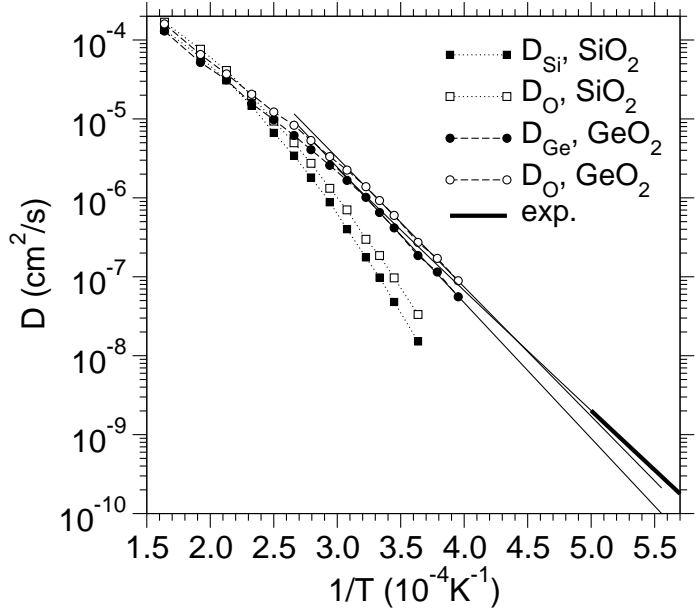}

\caption{\label{fig12} Self-diffusion constants of Ge and O in
GeO$_2$ melts plotted vs.~inverse temperature. For comparison,
results for Si and O in SiO$_2$ melts (taken from \cite{23}) are
included. Straight lines indicate fits to the Arrhenius relation,
Eq.~(\ref{eq11}). Also an Arrhenius fit resulting from
experimental viscosity data \cite{35} via Eq.~(\ref{eq13}) is
included.}
\end{center}
\end{figure}
In Fig.~\ref{fig12} a plot of the diffusion constants is presented,
choosing a logarithmic ordinate scale and inverse temperature as
abscissa, so Arrhenius relations show up via straight lines, since then
[compare to Eq.~(\ref{eq1})]
\begin{equation} \label{eq11}
D_\alpha=D_{\alpha, \infty} \, \exp [-E_{{\rm a}, \alpha}/(k_BT)] \, .
\end{equation}
The activation energies resulting from the fits in Fig.~\ref{fig12}
are $E_{\rm a, Ge}=3.41$\,eV and $E_{\rm a, O}=3.25$\,eV. As can be also
infered from Fig.~\ref{fig12}, oxygen diffuses slightly faster than Ge,
and this difference becomes slightly more pronounced with decreasing
temperature, due to the slightly higher activation energy of Ge. A
similar behavior is well-known for SiO$_2$ \cite{6,23}.

\begin{figure}
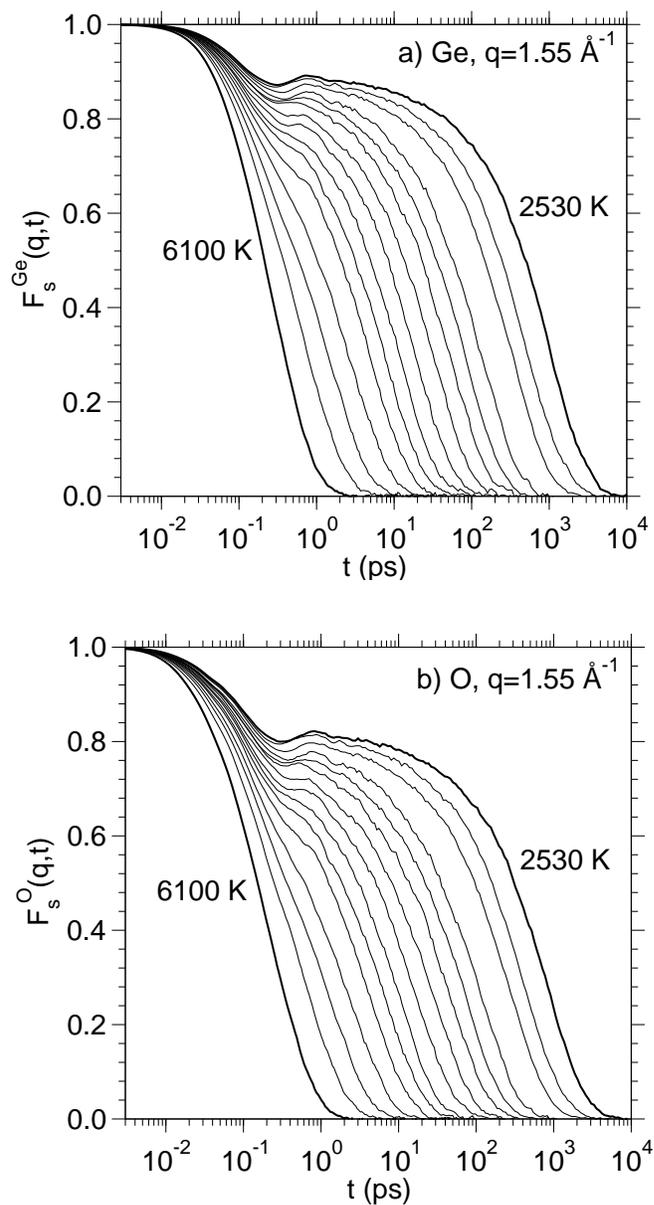

\begin{center}
\vspace*{0.5cm}
\includegraphics[width=0.65\textwidth]{fig13a.eps}
\vspace*{0.6cm}

\includegraphics[width=0.65\textwidth]{fig13b.eps}

\caption{\label{fig13} Incoherent intermediate scattering
functions $F_{\rm s}^{\rm Ge}(q,t)$ (a) and $F_{\rm s}^{\rm O} (q,t)$ 
(b) for GeO$_2$ at $q=1.55$\,\AA$^{-1}$, where the static structure 
factor $S(q)$ exhibits the first sharp diffraction peak, plotted 
vs.~time (on logarithmic scale) for a broad range of temperatures (the leftmost
curve corresponds to $T=6100$\,K, the rightmost curve to $T=2530$\,K,
temperatures in between are 5200\,K, 4700\,K, 4300\,K, 4000\,K, 3760\,K,
3580\,K, 3400\,K, 3250\,K, 3100\,K, 3000\,K, 2900\,K, 2750\,K, and 2640\,K).}
\end{center}
\end{figure}
While in the case of SiO$_2$ experimental data for self-diffusion
constants $D_{{\rm Si}}$, $D_{{\rm O}}$ are available, we are not
aware of suitable data for GeO$_2$. However, when we disregard the
small difference between $D_{{\rm Ge}}$ and $D_{\rm O}$, a rough
estimation of these diffusion constants is possible with the well-known
Stokes-Einstein relation \cite{6,73},
\begin{equation} \label{eq13}
D=k_BT / [c \pi \eta R] \, ,
\end{equation}
with the constant $c=4$ if one assumes slip boundary conditions for
the particle diffusion in the fluid, $\eta$ is the shear viscosity of
the fluid, and $R$ the radius of the diffusing particle. In principle,
Eq.~(\ref{eq13}) is a result from hydrodynamics, and makes sense only
if $R$ is much larger than interatomic distances. However, in the
spirit of the finding that often descriptions based on hydrodynamics
work down to the molecular scale (see \cite{76} for a recent example),
Eq.~(\ref{eq13}) is used also for diffusing atoms or molecules. Using
then for $R$ the Ge-O nearest neighbor distance, $R=1.75$~\AA, the
experimental viscosity data of Riebling \cite{35} are readily converted
into the self-diffusion constant, and the resulting Arrhenius fit
(implying $E_{\rm a}=3.565$\,eV) is also included in Fig.~\ref{fig12}
and in very good agreement with our simulations.

In contrast to our results, previous simulations \cite{48,49} gave
activation energies in the range between 1\,eV and 1.2\,eV. There are
many indications that the potential used by Hoang \cite{48} cannot
describe GeO$_2$ as accurately as the OE potential does; moreover
his results presumably suffer from aging effects due to insufficient
equilibration. The latter criticism also applies to the study of
Micoulaut {\it et al} \cite{49}, where the system configurations were
taken from one cooling run applying a cooling rate of $2.5 \times
10^{12}$\,K/s, although states in the temperature range from 940\,K $\leq
T \leq 2480$\,K were considered. It is clear that such configurations are
far from equilibrium, even at $T=2940$\,K these data \cite{49} do not show
any sign of the cage effect, and at the lower temperature ($T=940$\,K),
which is only about 100\,K higher than the experimental glass transition
temperature, the structural relaxation time is only of the order of ns,
which proves that the melt is in a state very far from equilibrium,
in the initial stages of aging.

We now turn to the analysis of intermediate scattering functions
(Fig.~\ref{fig13}). Again we note the qualitative similarity of these
curves to data for many other glassforming fluids \cite{6,23}. While
at high temperatures the decay of $F_{\rm s}^{\alpha}(q,t)$ resembles
a simple exponential, for $T \leq 3400$\,K the decay occurs in two
steps, due to the cage effect. The so-called ``$\beta$-relaxation''
is the time regime around the plateau, while the final decay from
this plateau to zero is called ``$\alpha$-relaxation'' \cite{3,6}.
Whereas, at high temperatures $F_{\rm s}^{\alpha}(q,t)$ decays to
zero on the ps timescale, at the lowest accessible temperatures the
``lifetime'' of the plateau extends into the ns time range already. In
order to define the structural relaxation time $\tau_{\alpha}(q,t)$,
we follow Ref.~\cite{30} by requesting that for $t=\tau_{\alpha}(q,t)$
the scattering function has decayed to a value of 0.1. Thus,
\begin{equation} \label{eq14}
F_{\rm s}^{\alpha}(q,t=\tau_{\alpha}(T)) =0.1 \, , \quad 
\alpha = {\rm Ge, O};
\end{equation}
here we have omitted the argument of the structural relaxation time,
since in the present context only the value of $q$, corresponding to
the location of the FSDP in the static structure factor, is of interest.

\begin{figure}
\begin{center}
\vspace*{0.5cm}
\includegraphics[width=0.7\textwidth]{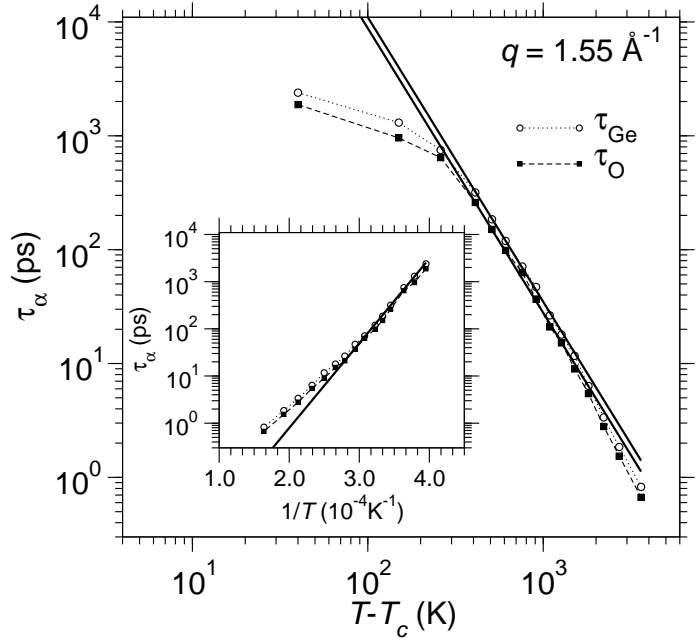}

\caption{\label{fig14} Log-log plot of $\tau_{\rm Ge}$ and 
$\tau_{\rm O}(T)$ in fluid GeO$_2$ versus $T-T_c$, using $T_c=2490$\,K. 
The bold lines are fits with power laws $\propto (T-T_c)^{\gamma}$
using $\gamma=2.5$. The insert shows the same data as 
${\rm log}[\tau_{\alpha}(T)]$ versus $1/T$. The straight line
in the insert is a fit with an Arrhenius law 
($\propto \exp[E_{\rm a}/(k_BT)]$) with $E_{\rm a}=3.57$\,eV.}
\end{center}
\end{figure}
Figure~\ref{fig14} presents a log-log plot of $\tau_{\rm Ge} (T)$ and
$\tau_{\rm O}(T)$ against $T-T_c$, $T_c=2490$\,K$\pm 100$\,K being an
estimate for the mode coupling \cite{3} critical temperature. Of course,
it is well-known that no real divergence of $\tau_{\alpha}(T)$, implying
an ergodic-to-nonergodic transition and a divergence of the viscosity at
$T_c$, can occur in real glassforming fluids \cite{6}: rather idealized
mode coupling theory \cite{3} is a kind of mean field theory for the
dynamic correlation functions of glassforming fluids, which supposedly
holds for $T>T_c$ but not too close to $T_c$, since the predicted
divergence at $T_c$ is rounded off. The standard albeit heuristic
interpretation is that an infinite lifetime of the cage ``imprisoning''
of the particles is prevented by thermally activated processes, so-called
hopping processes, which break up the cage even for $T=T_c$ and
$T<T_c$. So $T_c$ only plays the role of a crossover temperature,
where the temperature dependence of $\tau_{\alpha}(T)$ crosses over
from the power law $\tau_\alpha (T) \propto (T-T_c)^{-\gamma}$ to an
Arrhenius law. Note that we find the value $\gamma=2.5$ for the critical
exponent. This value is slightly higher than the one estimated from
simulations of silica \cite{30}. The insert of Fig.~\ref{fig14} shows
that indeed at lower temperatures our estimates for $\tau_\alpha (T)$
are consistent with an Arrhenius law. The activation energy is $E_{\rm
a}=3.57$\,eV in this case, i.e.~slightly higher than those determined
for the self-diffusion constants (see Fig.~\ref{fig12}). Of course,
this crossover from a power law to thermally activated behaviour is by no
means sharp, but actually rather gradual, and this smooth behaviour near
$T_c$ necessarily prevents us from an accurate estimation of $T_c$: data
near $T_c$ may deviate from the straight line on the log-log plot due to
the onset of the crossover, even if the estimate for $T_c$ is correct;
thus, the precise range of temperatures for which $\tau_{\alpha}(T)$
should be fitted to the power law is somewhat uncertain, and this leads
to a considerable uncertainty about $T_c$.

\begin{figure}
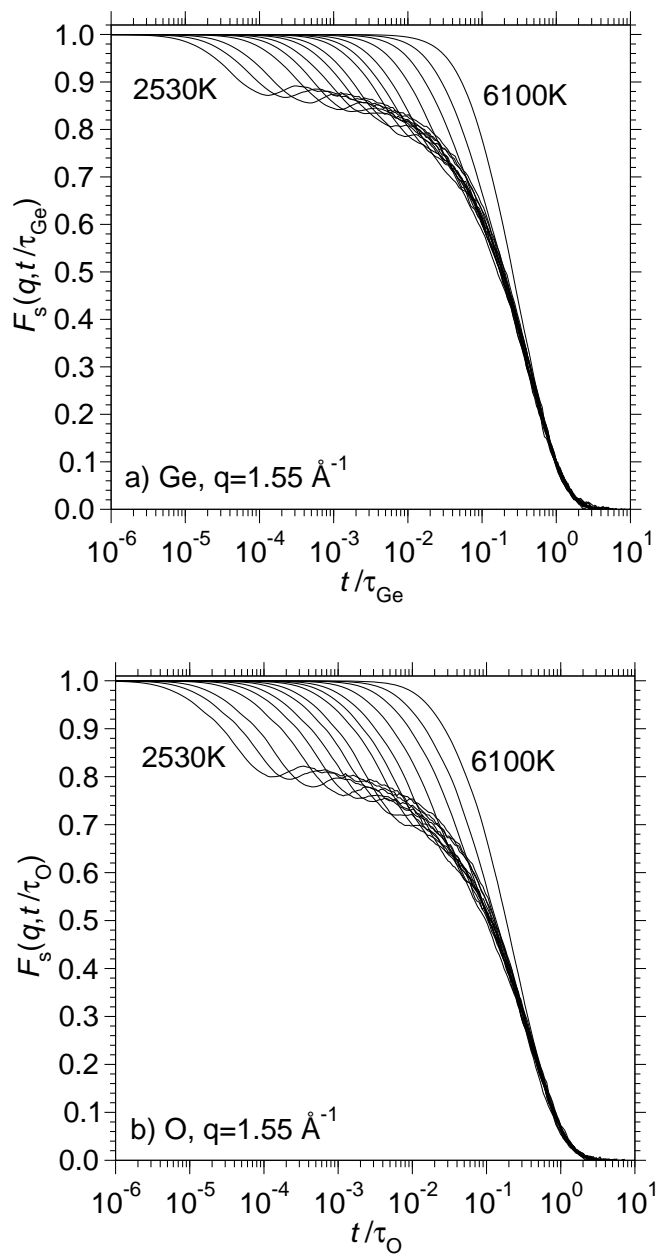

\begin{center}
\vspace*{0.5cm}
\includegraphics[width=0.65\textwidth]{fig15a.eps}
\vspace*{0.7cm}

\includegraphics[width=0.65\textwidth]{fig15b.eps}

\caption{\label{fig15} Incoherent intermediate scattering functions
$F_{\rm s}^{\alpha}(q,t)$ plotted vs.~the scaled time
$t/\tau_{\alpha}$ for Ge (a) and O (b). The temperatures shown are the
same as those of Fig.~\ref{fig13}.}
\end{center}
\end{figure}
As further evidence for the applicability of mode coupling theory
to describe the dynamics of molten GeO$_2$ at high temperatures,
Fig.~\ref{fig15} shows a test of the time temperature superposition
principle \cite{3,6}. Of course, the $\beta$-relaxation is not supposed
to follow this $\alpha$-scaling and thus the upper part of the curves
splay out, while the decay of the plateau (for temperatures where
a plateau exist) follow this scaling nicely. The quality with which
this scaling holds clearly is comparable to that observed for silica
and fragile glassformers.

In SiO$_2$, Saika-Voivod {\it et al} \cite{29} have tried to link the
relaxation dynamics to structural anomalies, such as the occurrence of
a density maximum \cite{29}. These authors have found some evidence for
the occurrence of a kind of liquid-liquid phase transition in fluid
SiO$_2$ at suitable conditions of temperature and pressure, i.e.~there
should exist two phases of fluid SiO$_2$ with different densities and
different structure of the random network of covalent bonds. If this
interpretation is correct, Figs.~\ref{fig12} and \ref{fig14} would
suggest that one should seek a similar interpretation in molten GeO$_2$,
too. However, our data (and the experimental data) for the density of
GeO$_2$ do not give any hint for structural anomalies of molten GeO$_2$
similar to those of SiO$_2$ (see Fig.~\ref{fig1}).

\begin{figure}
\begin{center}
\vspace*{0.5cm}
\includegraphics[width=0.7\textwidth]{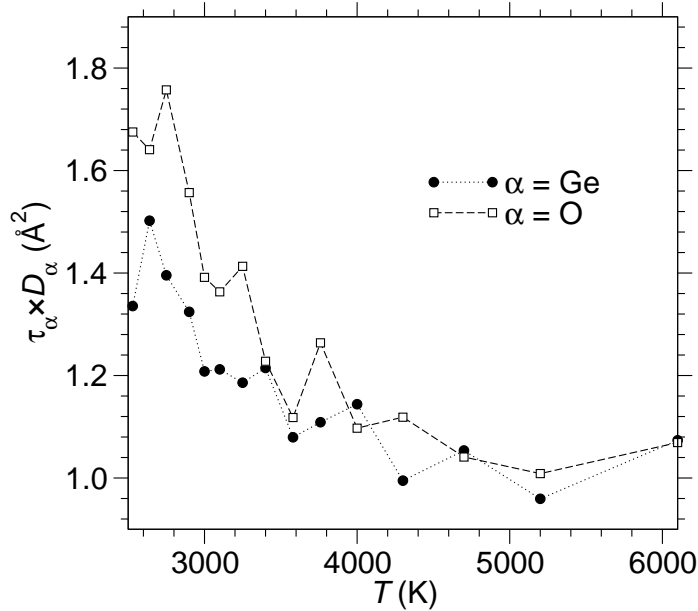}

\caption{\label{fig16} Plot of the product $\tau_{\alpha}(T)D_{\alpha}(T)$ 
vs.~temperature, for $\alpha=Ge$ and $\alpha=O$, as indicated. Simulation 
results (data points) are connected by lines to guide the eye.}
\end{center}
\end{figure}
Finally, we address the question to what extent the Stokes-Einstein
relation is valid for germania. In Eq.~(\ref{eq13}), this relation was
used to link experimental viscosity data \cite{35} to the self-diffusion
constants (Fig.~\ref{fig12}). While it would be possible to estimate
the shear viscosity from the time-correlation of the off-diagonal
pressure tensor components \cite{51,52}, and such an approach has been
shown feasible also for molten SiO$_2$ \cite{23}, the statistical
errors of the resulting estimates are very large. Since one usually
associates the shear viscosity with the structural relaxation time,
we tentatively tested to what extent the product $\tau_{\alpha}(T)
D_{\alpha}(T)$ is constant (Fig.~\ref{fig16}). Similar as in SiO$_2$,
where the product $\eta(T)D_{\alpha}(T)/T$ could be studied \cite{23},
one sees some increase of this Stokes-Einstein ratio, but the increase
again is not really dramatic. Much more dramatic violations of the
Stokes-Einstein relation have been observed for various materials near
$T_g$, and this has found a lot of attention in the literature (see
e.g.~\cite{14,77,78,79}). All of our data are far above the melting
temperature of GeO$_2$, and we can say that in this temperature regime a
dramatic breakdown of the Stokes-Einstein relation does not occur. If a
breakdown occurs near $T_g$, this would imply that the good agreement
between the experimental data in Fig.~\ref{fig12} and our estimates
for the self-diffusion constants is merely accidental.

\section{Conclusions}

In the present work, we have described the results of a simulation study
of fluid GeO$_2$ at zero pressure, based on extensive MD runs using the
Oeffner-Elliott potential. In the temperature region $T\geq 2530$\,K
the melt has been carefully equilibrated, while glassy structures of
amorphous GeO$_2$ at room temperature were also produced, cooling
down the system with two different cooling rates which did produce
only minor structural differences, however. To validate our potential,
also an ``ab initio'' Car-Parrinello Molecular Dynamics (CPMD) study was
performed, which is limited to even higher temperatures ($T\geq3000$\,K)
and very small systems ($N\leq120$ atoms). When we consider
properties at small enough scales where systematic errors due to finite
size do not matter, we find very good agreement between the MD and CPMD
descriptions, for both static and dynamics properties. The most important
distinction is that the effective potential to which CPMD corresponds
is slightly softer than the OE potential. However, the OE potential
is clearly more accurate than other (empirical) potentials that were
used in the literature, and our results agree also rather well with the
(albeit somewhat scarce) experimental data which are available so far.

Having validated the OE potential, it would be interesting to use it
both for a more complete study of the dynamics of molten and glassy
GeO$_2$, and for a careful study of GeO$_2$ under pressure. This must be left
to future work. Also it would be interesting to use CPMD to construct
a new effective potential which is even more accurate than the OE
potential. Such attempts have been made \cite{59}, but so far were
not successful.

Our results suggest that for $T \geq 2500$\,K a crossover sets in
for the dynamical properties from a thermally activated behaviour of
various quantities to a behaviour described by mode coupling theory,
similar to previous findings for silicon dioxide. In view of our results
it is not too surprising that in the temperature range $T\leq1600$\,K
no experimental evidence for mode coupling effects could be found
\cite{41}. Thus, it would be very useful if measurements could be
extended to higher temperatures. Also a measurement of self-diffusion
constants would be useful, to test the finding that only a rather weak
violation of the Stokes-Einstein relation occurs in GeO$_2$.

GeO$_2$ and SiO$_2$ are the two archetypical examples for strong
glassformers. Our results imply that they behave qualitatively similar,
including the crossover to mode coupling type behavior at temperatures
far above melting. The latter crossover is also characteristic for
fragile glassformers. However, it is interesting that, in contrast
to typical fragile glassformers, the critical mode coupling temperature 
is above the melting temperature. It has to be seen in future studies
whether the relative location of critical temperature and melting
temperature can be used as a criterion to distinguish fragile
and strong glassformers. 

In conclusion, we hope that the present work helps to sort out the many
questions concerning the possible universality (or lack thereof) in
glassforming fluids, and stimulate further experimental and theoretical
work on the above issues.

\ack{We thank Th. Voigtmann for a critical reading of the manuscript.
We are indebted to P. S. Salmon for providing the experimental data
included in the comparison in Fig.~2.
One of us (M.H.) is grateful to the SCHOTT AG (Mainz) for
financial support. We are grateful to U.~Fotheringham, W.~Kob and
M.~Letz for stimulating discussions, and acknowledge a substantial
grant of computer time at the J\"ulich multiprocessor system (JUMP)
of the John von Neumann Institute for Computing (NIC).}

\section*{References}


\begin{thebibliography}{99}
%
%
\bibitem{1} 
Zallen R 1983 
{\it The Physics of Amorphous Solids} (New York: Wiley)
%
\bibitem{2} 
J\"ackle J 1986 
{\it Rep. Prog. Phys.} {\bf 49} 171
%
\bibitem{3} 
G\"otze W and Sj\"ogren L 1992 
{\it Rep. Prog. Phys.} {\bf 55} 241
%
\bibitem{4} 
Debenedetti P G 1997 
{\it Metastable Liquids} (Princeton: Princeton Univ. Press)
%
\bibitem{5} 
Angell C A, Ngai K L, Kieffer J, Egami T and Nienhaus G U (eds.) 1997 
{\it Structure and Dyamics of Glasses and Glass Formers} 
(Pittsburg: Mat. Res. Soc.)
%
\bibitem{6} 
Binder K and Kob W 2005 
{\it Glassy Materials and Disordered Solids: An Introduction to Their 
Statistical Mechanics} (Singapore: World Scientific)
%
\bibitem{7} 
Andreozzi L, Giordano M., Leporini D and Tosi M (eds.) 2007
{\it Proceedings of the Fourth Workshop on Nonequilibrium
Phenomena in Supercooled Fluids, Glasses and Amorphous Materials,
Pisa, 17-22 September 2006} J. Phys.: Condens. Matter {\bf 19},
special issue
%
\bibitem{8} 
Fisher M E 1974 
{\it Rev. Mod. Phys.} {\bf 46} 597
%
\bibitem{9} 
Angell C A 1985 
in {\it Relaxations in Complex Systems}
eds Ngai K L and Wright G B (Springfield: US Dept. of Commerce)
%
\bibitem{10} 
G\"otze W 1999 
{\it J. Phys.: Condens. Matter} {\bf 11} A1
%
\bibitem{11} 
Biroli G and Bouchaud J-P 2007 
{\it J. Phys.: Condens. Matter} {\bf 19} 205101
%
\bibitem{12} 
Franz S and Parisi G 2000 
{\it J. Phys.: Condens. Matter} {\bf 12} 6335
%
\bibitem{13} 
Eastwood M P and Wolynes P G 2002 
{\it Europhys. Lett.} {\bf 60} 587
%
\bibitem{14} 
Jung Y, Garrahan J P and Chandler D 2004 
{\it Phys. Rev. E} {\bf69} 061205
%
\bibitem{23}
Horbach J and Kob W 1999
{\it Phys. Rev. B} {\bf 60} 3169
%
\bibitem{27}
Sciortino F and Kob W 2001
{\it Phys. Rev. Lett.} {\bf 86} 648
%
\bibitem{30}
Horbach J and Kob W 2001
{\it Phys. Rev. E} {\bf 64} 041503
%
\bibitem{15} 
Angell C A, Clarke J H R and Woodcock L V 1981 
{\it Adv. Chem. Phys.} {\bf 48} 397
%
\bibitem{16} 
Rustad J R, Yuen D A and Spera F J 1990 
{\it Phys. Rev. A} {\bf 42} 2081
%
\bibitem{17} 
Dellavalle R G and Venuti E 1994 
{\it Chem. Phys.} {\bf 179} 411; 
{\it ibid} 1996 {\it Phys. Rev. B} {\bf 54} 3809
%
\bibitem{18} 
Tsuneyuki S and Matsui Y 1995 
{\it Phys. Rev. Lett.} {\bf 74} 3197
%
\bibitem{19} 
Vollmayr K, Kob W and Binder K 1996 
{\it Phys. Rev. B} {\bf 54} 15808
%
\bibitem{20} 
Horbach J, Kob W, Binder K and Angell C A 1996 
{\it Phys. Rev. E} {\bf 54} R5897
%
\bibitem{21} 
Pasquarello A and Car R 1997 
{\it Phys. Rev. Lett.} {\bf 79} 1766
%
\bibitem{22} 
Horbach J, Kob W and Binder K 1998 
{\it J. Non-Cryst. Solids} {\bf 235-238} 320
%
\bibitem{24} 
Horbach J, Kob W and Binder K 1999 
{\it J. Phys. Chem. B} {\bf 103} 4104
%
\bibitem{25} 
Jund P and Jullien R 1999 
{\it Phys. Rev. Lett.} {\bf 83} 2210
%
\bibitem{26} 
Benoit M, Ispas S, Jund P and Jullien R 2000 
{\it Eur. Phys. J. B} {\bf 13} 631
%
\bibitem{28} 
Horbach J, Kob W and Binder K 2001 
{\it Eur. Phys. J. B} {\bf 19} 531
%
\bibitem{29} 
Saika-Voivod I, Sciortino F and Poole P H 2001 
{\it Phys. Rev. E} {\bf 63} 011202
%
\bibitem{31} 
Shell M S, Debenedetti P G and Panagiotopoulos A Z 2002
{\it Phys. Rev. E} {\bf 66} 011202
%
\bibitem{32} 
Vogel M and Glotzer S C 2004 
{\it Phys. Rev. Lett.} {\bf 92} 255901
%
\bibitem{33} 
Takada A, Richet P, Catlow C R A and Price G D 2004
{\it J. Non-Cryst. Solids} {\bf 345-346} 224
%
\bibitem{34} 
Huang L and Kieffer J 2004 
{\it Phys. Rev. B} {\bf 69} 224203
%
\bibitem{35} 
Riebling E F 1963 
{\it J. Chem. Phys.} {\bf 39} 3022
%
\bibitem{36} 
Fontana E H and Plummer W A 1966 
{\it Phys. Chem. Glasses} {\bf 7} 139
%
\bibitem{37} 
Bondot P 1974 
{\it Acta Cryst. A} {\bf30} 470;
Desa J A E, Wright A C and Sinclair R N 1988 
{\it J. Non-Cryst. Solids} {\bf 99} 276; 
Waseda Y, Sugiyama K, Matsubara E and Harada K 1990
{\it Mat. Trans. JIM} {\bf 31} 421
%
\bibitem{38} 
Dingwell D B, Knoche R and Webb S L 1993 
{\it Phys. Chem. Min.} {\bf 19} 445
%
\bibitem{39} 
Price D L, Saboungi M-L and Barnes A C 1998 
{\it Phys. Rev. Lett.} {\bf 81} 3207
%
\bibitem{40} 
Price D L, Ellison A J G, Sabougni M-L, Hu R-Z, Egami T and Howells WS 1997 
{\it Phys. Rev. B} {\bf 55} 11249;
Neuefeind J and Liss K-D 1996 
{\it Ber. Bunsenges. Phys. Chem.} {\bf 100} 1341
%
\bibitem{41} 
Meyer A, Schober H and Neuhaus J 2001 
{\it Phys. Rev. B} {\bf 63} 212202
%
\bibitem{42} 
Sampath S, Benmore C J, Lantzky K M, Neuefeind J, Leinenweber K, 
Price D L and Yarger J L 2003 
{\it Phys. Rev. Lett.} {\bf 90} 115502
%
\bibitem{43} 
Ohtaka O, Arima H, Fukui H, Utsumi W, Katayama Y and Yoshiasa A 2004 
{\it Phys. Rev. Lett.} {\bf 92} 155506
%
\bibitem{salmon06}
Salmon P S, Barnes A C, Martin R A and Cuello G J 2006
{\it Phys. Rev. Lett.} {\bf 96} 235502
%
\bibitem{salmon07_1}
Salmon P S, Barnes A C, Martin R A and Cuello G J 2007
{\it J. Phys.: Condens. Matter} {\bf 19} 415110
%
\bibitem{salmon07_2}
Salmon P S 2007
{\it J. Phys.: Condens. Matter} {\bf 19} 455208
%
\bibitem{44} 
Oeffner R D and Elliott S R 1998 
{\it Phys. Rev. B} {\bf 58} 14791
%
\bibitem{45} 
Gutierrez G and Rogan J 2004 
{\it Phys. Rev. E} {\bf 69} 31201
%
\bibitem{46} 
Micoulaut M 2004 
{\it J. Phys.: Condens. Matter} {\bf 16} L131
%
\bibitem{47} 
Giacomazzi L, Umari P and Pasquarello A 2005 
{\it Phys. Rev. Lett.} {\bf 95} 075505
%
\bibitem{48} 
Hoang V V 2006 
{\it J. Phys.: Condens. Matter} {\bf 18} 777
%
\bibitem{49} 
Micoulaut M, Guissani Y and Guillot B 2006 
{\it Phys. Rev. E} {\bf 73} 031504
%
\bibitem{50} 
Shanavas K V, Garg N and Sharma S M 2006 
{\it Phys. Rev. B} {\bf 73} 094120
%
\bibitem{giaccomazzi06}
Giacomazzi L, Umari P and Pasquarello A 2006
{\it Phys. Rev. B} {\bf 74} 155208
%
\bibitem{micoulaut06_2}
Micoulaut M, Cormier L and Henderson G S 2006 
{\it J. Phys.: Condens. Matter} {\bf 18} R753
%
\bibitem{51} 
Allen M P and Tildesley D J 1987 
{\it Computer Simulation of Liquids} (Oxford: Clarendon Press)
%
\bibitem{52} 
Binder K and Ciccotti G (eds) 1996 
{\it Monte Carlo and Molecular Dynamics of Condensed Matter} 
(Bologna: Societa Italiana de Fisica)
%
\bibitem{53} 
Car R and Parrinello M 1985 
{\it Phys. Rev. Lett.} {\bf 55} 2471
%
\bibitem{54} Marx D and Hutter J 2000 
in {\it Modern Methods and Algorithms of Quantum Chemistry}
ed Grotendorst J (J\"ulich: NIC)
%
\bibitem{55} 
http://www.cpmd.org/
%
\bibitem{56} 
van Beest B H W, Kramer G J and van Santen R A 1990 
{\it Phys. Rev. Lett.} {\bf 64} 1955
%
\bibitem{57} 
Tsuchya T, Yamanaka T and Matsui M 1998 
{\it Phys. Chem. Minerals} {\bf 25} 94
%
\bibitem{58} 
Karthikeyan A and Almeida R M 2001 
{\it J. Non-Cryst. Solids} {\bf 281} 152
%
\bibitem{59} 
Hawlitzky M 2006 
{\it Dissertation} (Mainz: Johannes Gutenberg Universit\"at)
%
\bibitem{hawlitzky07}
Hawlitzky M, Horbach J and Binder K 2007
{\it MRS Proc. Symp.} {\bf 1048E} Z9.1
%
\bibitem{60} 
Andersen H C 1980 
{\it J. Chem. Phys.} {\bf 72} 2384
%
\bibitem{62} 
Goedecker S, Teter M and Hutter J 1996 
{\it Phys. Rev. B} {\bf 54} 1703
%
\bibitem{63} 
Troullier N and Martins J L 1991 
{\it Phys. Rev. B} {\bf 43} 1993
%
\bibitem{64} 
Martyna G J, Klein M L and Tuckerman M 1992 
{\it J. Chem. Phys.} {\bf 97} 2635
%
\bibitem{tuckerman94}
Tuckerman M E and Parrinello M 1994
{\it J. Chem. Phys.} {\bf 101} 1302
%
\bibitem{65} 
Mischler C, Kob W and Binder K 2002 
{\it Computer Phys. Comm.} {\bf 147} 222
%
\bibitem{66} 
Mischler C, Horbach J, Kob W and Binder K 2005 
{\it J. Phys.: Condens. Matter} {\bf 17} 4005
%
\bibitem{67} 
Sarver J F and Hummel F A 1960 
{\it J. Am. Ceram. Soc.} {\bf 43} 336
%
\bibitem{68} 
K\"uhne T D, Krack M, Mohamed F R and Parrinello M 2007
{\it Phys. Rev. Lett.} {\bf 98} 066401
%
\bibitem{69} 
Angell C A and Kanno H 1976 
{\it Science} {\bf 193} 1121
%
\bibitem{70} 
see {\it NIST Neutron Scattering lengths and Cross
Sections} available at
http://www.ncnr.rist.gov/resources/n-lengths/
%
\bibitem{71} 
Elliott S R 1992 
{\it J. Non-Cryst. Solids} {\bf 150} 112
%
\bibitem{72} 
Price D L and Carpenter J M 1987 
{\it J. Non.-Cryst. Solids} {\bf 92} 153
%
\bibitem{73} 
Hansen J P and McDonald I R 1990
{\it Theory of Simple Liquids} (London: Academic)
%
\bibitem{74} 
Tamura T, Lu G-H, Yamamoto R and Kohyama M 2004
{\it Phys. Rev. B} {\bf 69} 195204
%
\bibitem{75} 
Neuefeind J and Liss K-D 1996 
{\it Ber. Bunsenges. Phys. Chem.} {\bf 100} 1341
%
\bibitem{76} 
Dimitrov D I, Milchev A and Binder K 2007 
{\it Phys. Rev. Lett.} {\bf 99} 054501
%
\bibitem{77} 
Berthier L, Chandler D and Garrahan J P 2005 
{\it Europhys. Lett.} {\bf 69} 320
%
\bibitem{78} 
Pan A C, Garrahan J P and Chandler D 2005 
{\it Chem. Phys. Chem.} {\bf 6} 1783
%
\bibitem{79} 
Szamel G and Flenner E 2006 
{\it Phys. Rev. E} {\bf 73} 011504
%
\end{thebibliography}
\end{document}